\documentclass[pre,twocolumn,amsmath,showpacs,floatfix]{revtex4}
 \usepackage{graphicx}

\begin{document}
\title{Directed percolation with incubation times}

\author{Andrea Jim\'enez-Dalmaroni}
\altaffiliation{Present address:  Max Planck Institut f\"ur Physik komplexer Systeme,
 N\"othnitzer Str. 38, D-01187 Dresden, Germany}
\email{andrea@mpipks-dresden.mpg.de}
\affiliation{Rudolf Peierls Centre for Theoretical Physics, University of Oxford, 1 Keble Road, Oxford OX1 3NP, UK}

\begin{abstract}
We introduce a model for directed percolation with a long-range temporal diffusion, while the spatial diffusion is kept short ranged.
In an interpretation of directed percolation as an epidemic process, this non-Markovian modification  can be understood as \textit{incubation times}, which are distributed accordingly to a L\'evy distribution. We argue that the best approach to find the
effective action for this problem is through a generalization of the
Cardy-Sugar method, adding the non-Markovian
features into the geometrical properties of the lattice. We formulate a field theory for this problem
and renormalize it up to one loop in a
perturbative expansion. We solve the various technical difficulties that
the integrations possess by means of an asymptotic analysis of the
divergences. We show the absence of field renormalization at one-loop order,
and we argue that this would be the case to all orders in perturbation theory.
Consequently, in addition to the characteristic scaling relations of directed percolation, we find a scaling relation valid for
the critical exponents of this theory. In this universality class, the critical
exponents vary continuously with the L\'evy parameter.
\end{abstract}

\maketitle
\def\xvec{{\text{\bf x}}}
\parskip 2mm

\section{Introduction}
\label{Introduction}

Directed percolation (DP) ~\cite{Kinzel83,HH2000,THV2005} is the generic model for nonequilibrium systems which exhibit a continuous phase
transition into a unique \textit{absorbing state}. DP describes the temporal-directed spreading of a
nonconserved agent in a certain medium. The agent might be bacteria in a
case of epidemics in populations, fire in a burning forest, or water in a porous rock. The spreading phenomenon is 
characterized by two competing processes: relying on the medium conditions, the agent may multiply itself or decay at a constant rate. 
Depending on the balance between these two processes, the spreading may continue forever or die out after certain time.
 If the agent is not allowed to appear spontaneously,
in the latter case the system is trapped in the absorbing state, a configuration where the stochastic 
fluctuations cease entirely and the system cannot escape from. These two regimes of survival and extinction are 
typically separated by a continuous phase transition characterized by the DP critical exponents.
 
The DP universality class is extremely robust, as a whole range of theoretical models seems to belong to
it. Some examples include heterogeneous catalysis~\cite{ZGB86}, chemical reactions~\cite{GdelaT79,Schlogl72}, 
interface depinning~\cite{TangLeschhorn92, Buldyrev}, the onset of
spatiotemporal chaos~\cite{RRR03}, flowing sand~\cite{FlowingSand} and self-organized 
criticality~\cite{MohantyDhar}. The robustness of the model has led to the conjecture that two-state spreading processes with
short-range interactions generically belong to the DP class, provided that
quenched randomness, unconventional symmetries, and large scales due to
memory effects are absent~\cite{Janssen81,Grassberger82}. 

In the context of epidemics, DP describes infection processes without immunization and where the disease is only 
transmitted to nearest neighbors by direct contact. This can be understood 
realizing the problem of DP on a $(d+1)$-dimensional lattice, where each lattice site
is considered as an individual which can be \textit{infected} (active) with
probability $p$ or \textit{healthy} (inactive) with probability $1-p$. 
An infected individual recovers at the next time step with
probability 1 and is ready to be reinfected with the same constant
probability $p$. The susceptibility to infection is then
independent of previous infections, and this ensures the absence of
immunization.

Therefore, in order to make a realistic description of epidemics the effect of immunization 
as well as long-range interactions should be taken into account as modifications of the DP model.
Immunization can be added to DP by considering a probability for
subsequent infections different from the first infection probability ~\cite{Cardy83}. This non-Markovian feature changes the universality class 
of the model to the one corresponding to dynamical percolation, also known as a
general epidemic process (GEP)~\cite{CardyGrassberger85,Janssen85}. The phase diagram of this model displays a curve phase transition line connecting the 
GEP and DP critical points~\cite{GCR97}. Along this line the same universality class as
GEP is found. A horizontal phase transition line also separates the GEP phase from a supercritical DP behavior.
The absence of scaling along this line has been shown for the case of $1+1$ dimensions~\cite{JDH03}
and later generalized to $d+1$ dimensions~\cite{JDthesis,DH2004}.

On the other hand, an epidemic model with long-range interactions was first suggested by Mollison~\cite{Mollison77}. This
model was studied as a generalization of DP which includes spatial long-range
interactions where the spreading distances follow a power-law distribution given by,
\begin{equation}
\label{LevyFlights}
\mathcal{P}(r) dr \sim r^{-d-\sigma} dr\,,
\end{equation}
where $d$ is the spatial dimension and $\sigma$ is a control
parameter. Asymptotically, as $r\rightarrow \infty$, Eq.~(\ref{LevyFlights}) is equivalent to a  L\'evy
distribution and $\sigma$ is called the  L\'evy exponent. In this
sense we can say that the particles perform L\'evy flights~\cite{Bouchaud90}.
The claim that the critical exponents should vary continuously with
$\sigma$~\cite{Grass86} was confirmed by theoretical renormalization-group analysis~\cite{Janssen99} as
well as by extensive numerical
simulations~\cite{Marques94,HHoward99}. More recently these results have
been generalized to branching-annihilating  L\'evy
flights~\cite{VernonHoward01} and to the pair annihilation reaction $A+A \to 0$~\cite{Vernon03}.
So far all these studies have assumed dynamic processes which are local
in time. In the case of epidemics, it is assumed that the infection happens instantaneously in time.

In order to make one step forward in making a more realistic
model for epidemics, in this paper we set out to study a non-Markovian modification of
the DP problem which includes waiting times, or \textit{incubation times}, between the infection and actual outbreak of the
disease in a population. We assume that these incubation
times $\tau$ are distributed  asymptotically as  $\tau\to \infty$
\begin{equation}
\label{Levydist_tau}
\mathcal{F}(\tau) d\tau \propto \frac{1}{\tau^{1+\kappa}} d\tau.
\end{equation}
Here the L\'evy parameter $\kappa>0$ is a free parameter that controls the characteristic
shape of the distribution. For $\tau\to 0$ we assume $\mathcal{F}(\tau)$ is a smooth
function of $\tau$. For simplicity we also assume that the dynamic
processes are local in space, which means that
the infection can only spread by contact with nearest neighbors.

The characterization of the universality class of this
problem has remained an open problem in the literature since it was
first suggested in a previous work~\cite{HHoward99} in 1999. This is mainly
because of the technical difficulties that arise in the field-theoretical description when long-range waiting times are introduced.
In the present work, we derive a field theory for this problem
and calculate the critical exponents by means of systematic
perturbation theory and the $\epsilon$ expansion. This paper is
organized as follows. In Sec.~\ref{DPlevyTheModel} we
propose a convenient approach to derive the field-theoretical action,
through a generalization of the method introduced by Cardy and Sugar
in~\cite{CardySugar80}. We dedicate Sec.~\ref{DPlevyMeanField} to study and
analyze the mean-field predictions of our theory. Subsequently in Sec.~\ref{DPlevyFieldTheory}
fluctuation effects are taken into account via renormalization-group methods.
The various difficulties that emerge through the renormalization process are managed
by studying the asymptotic behaviour of the integrals involved in
the renormalization process. Finally, in
Sec.~\ref{DPlevyCallan} we write the renormalization group
equations and compute the critical exponents at one loop. In
Sec.~\ref{DPlevyDiscussion}, devoted to the conclusion, we argue that our
results are valid to any loop order.

\section{The Model}
\label{DPlevyTheModel}

\subsection{Master Equation}
\label{themasterequation}

In order to derive the field-theoretical action for the
problem of DP with incubation times, we first consider the master equation
formalism. Directed percolation can be interpreted as a reaction-diffusion
process of identical particles in a $d$-dimensional lattice, where multiple occupation
is allowed. We call $P(\alpha, t)$ the probability that the system will be at a given
microstate $\alpha$. The dynamics of such a system usually is described by
a master equation governing the temporal evolution of the probability distribution
$P(\alpha,t)$, which its general form is given by
\begin{equation}
\label{GeneralMasterEquation}
\frac{dP(\alpha,t)}{dt}=\sum_{\beta}R_{\beta\rightarrow\alpha}P(\beta,t)-\sum_{\beta}R_{\alpha\rightarrow\beta}P(\alpha,t).
\end{equation}
The system goes from the microstate $\alpha$ to the microstate $\beta$
with a constant transition rate $R_{\alpha\rightarrow\beta}$.
A naive generalization of this to processes involving transitions with
incubation times would be 
\begin{multline}
\label{GeneralMasterEquation2}
\frac{dP(\alpha,t)}{dt} \stackrel{?}{=} \sum_{\beta} \int_{t'<t} dt'
R_{\beta\rightarrow\alpha}(t-t')P(\beta,t') \, - \\ 
\sum_{\beta} \int_{t<t''} dt'' R_{\alpha\rightarrow\beta}(t''-t)P(\alpha,t),
\end{multline}
where the transition rates $R_{\alpha\rightarrow\beta}$ are time-dependent functions. But Eq.~(\ref{GeneralMasterEquation2}) is wrong, as the
probabilities $P(\beta,t')$ do not refer to mutually exclusive events for
different times $t'$. Indeed,
Eq.~(\ref{GeneralMasterEquation2}) does not conserve the total probability 
 $\sum_{\alpha}P(\alpha,t)$ . In fact Eq.~(\ref{GeneralMasterEquation2})
describes the dynamics of particles, which disappear from the lattice at time $t'$,
until they reappear at time $t>t'$. This does not correspond to the dynamics
with incubation times that we are trying to model.  

To write a correct master equation one should add to
the right-hand side of Eq.~(\ref{GeneralMasterEquation2}) an
infinite number of terms which will take into account the nondisjoint  nature
of the events, so the master equation is replaced by an infinite hierarchy of
coupled equations for multitime joint probabilities.

In order to avoid dealing with the difficulties inherent to a master equation formalism,
we propose to adopt an alternative way to find the field-theoretical
action. We will generalize a method first introduced by Cardy and
Sugar in order to show that directed bond percolation was in the same universality class of
Reggeon field theory~\cite{CardySugar80}.

\subsection{Action of DP with incubation times}

\begin{figure}
\centering
\includegraphics[width=50mm]{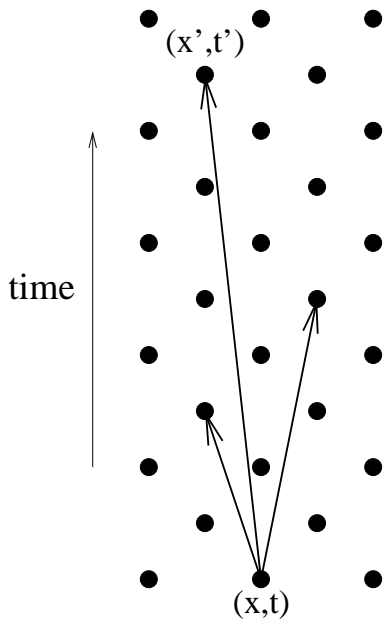}
\caption{
\label{inf_vectors}
Example on a (1+1)-dimensional lattice of possible temporal long-range
infection vectors from a lattice site $(x,t)$ to nearest-neighbor sites in space.
}
\end{figure}

In order to provide a model for epidemics with long incubation times, we
consider our system on a $(d+1)$-dimensional lattice. We represent the spreading
of the infection on the lattice through vectors (see Fig.~\ref{inf_vectors}). An infection vector between a
lattice site $(x,t)$ and another site $(x',t')$ is present with probability
$p(x'-x,t'-t)$, with $t<t'$. The temporal coordinate $t$ indicates the preferred direction, and
therefore the orientation of the infection vectors is always in the direction
of the increasing time. The vectors can only connect nearest neighbors in space, but their range
in time depends on the incubation times distributed as
Eq.~(\ref{Levydist_tau}). Considering this model, the problem of epidemics
with long incubation times can be interpreted as a temporal long-range directed
percolation problem. We propose now to write a field theory for this model,
through a generalization of the Cardy-Sugar method.
We define the connectivity function
$G(x_2,t_2;x_1,t_1)$ as the probability for two given lattice sites
$(x_1,t_1)$ and $(x_2,t_2)$, with $t_1<t_2$, to be connected
irrespective of the other sites. We say that two sites are connected if there
is an infection vector present between them. Following Cardy and
Sugar~\cite{CardySugar80}, $G$ can be written as
\begin{multline}
\label{Gandcentralfactor}
G(x_2,t_2;x_1,t_1)= \mbox{Tr}\,a(x_2,t_2)\prod_{links, t^\prime>t}[1+p(x^\prime-x,t^\prime \\
-t)\bar{a}(x^\prime,t^\prime) a(x,t)] \bar{a}(x_1,t_1). 
\end{multline}
The commuting operators $a(x,t)$ and $\bar{a}(x,t)$ act on each site
$(x,t)$ of the lattice, and their algebra is defined as~\cite{CardySugar80}
\begin{equation}
a^2=ia \mbox, \, \,\, \, \,\,\, \,\,\bar{a}^2=i \bar{a},
\end{equation}
\begin{equation}
\mbox{Tr} a(x,t)= \mbox{Tr}\bar{a}(x,t)=0,
\end{equation}
\begin{equation}
\mbox{Tr}[a(x,t) \bar{a}(x,t)]=1.
\end{equation}
We remark that Eq.~(\ref{Gandcentralfactor}) is identical to the one obtained by
Cardy and Sugar in ~\cite{CardySugar80}, except for the fact that in our case the probability $p$ is
not a constant.

The physical features of the problem and the details of the dynamics are
included in the effective lattice determined by the infection
vectors. We define a matrix $V$, which will contain this information
as follows:
\begin{multline}
\label{V_def}
\prod_{links, t^\prime>t}[1+p(x'-x,t'-t)\bar{a}(x^\prime,t^\prime) a(x,t)] \\
=\exp\biggl(\sum_{x,t} \sum_{x^\prime,t^\prime} \bar{a}(x^\prime,t^\prime) V(x^\prime-x,t^\prime-t) a(x,t)\biggr).
\end{multline}
Therefore the matrix $V$ will contain the information of the temporal long-range
processes. In order to complete the generalization
of the Cardy-Sugar method for the problem of DP with incubation times, we assume that $V$ can be decomposed into a short-range part
$V_{s}(x,t)$ and other part $V_{l}(x,t)$ which will be long range in time,
\begin{equation}
V(x,t)=V_{s}(x,t)+V_{l}(x,t).
\end{equation}
$V_{l}(x,t)$ contains the long-range temporal dependence and a factor with a spatial
dependence, which is short ranged. Therefore, we can assume that the leading behavior of $V_{l}(x,t)$ is as follows:
\begin{equation}
V_{l}(x,t) \sim \frac{1}{t^{1+\kappa}},
\end{equation}
with a proportionality factor that is some short-range function of $x$.
We consider an expansion of the Fourier-Laplace transform of $V_{s}$ in a small momentum $k$ and a
small energy $E$,
\begin{equation}
\tilde{V}_{s}(k,E)=\sum_{x,t} \biggl(1-\frac{1}{2}(kx)^2-Et+\cdots\biggr)V_{s}(x,t),
\end{equation}
where
\begin{equation}
\tilde{V}_{s}(k,E)=c-c_1 E- c_2 k^2+\cdots.
\end{equation}
In the case of long-range temporal processes considered here, the moment $\langle t \rangle$
is divergent and we cannot perform an expansion in $E$.  
Instead, we compute the Laplace transform of $\mathcal{F}(t)$ as
\begin{equation}
\int_{0}^{\infty}e^{-Et} \mathcal{F}(t) dt \sim E^{\kappa} + \mbox{const}+
\mbox{regular terms}.
\end{equation}
Therefore, the Fourier-Laplace transform of the long-range contribution
$V_{l}(x,t)$ will involve a $E^{\kappa}$ dependence multiplied by the Fourier
expansion of a spatial short-range factor,
\begin{equation}
\tilde{V}_{l}(k,E) \sim (E^{\kappa}+\mbox{const}+ \cdots) (1- b k^2 +\cdots).
\end{equation}
We should notice that this is valid for values of $\kappa>0$.
Keeping the most relevant terms in a small-$k$ and -$E$ expansion,
the  Fourier-Laplace transform of $V$ is given by
\begin{equation}
\tilde{V}(k,E)=c[1-r_1 E - r_2 k^2 - r E^{\kappa} + O(k^2 E^{\kappa})].
\end{equation}
Applying  Gaussian integrations in Eq.~(\ref{V_def}), we can be written as
\begin{multline}
\label{fact_prod}
\exp\biggl(\sum_{x,t} \sum_{x^\prime,t^\prime} \bar{a}(x^\prime,t^\prime) V(x^\prime-x,t^\prime-t) a(x,t)\biggr)\\
=\int_{-\infty}^{\infty} \prod_{(x,t)} d\phi d\bar{\phi} \exp[\bar{\phi}(x^\prime,t^\prime)(V)^{-1} \phi(x,t)-a\bar{\phi}-\bar{a}\phi],
\end{multline}
where the operator $V^{-1}$ is given by
\begin{equation}
\label{Vmatrix}
V^{-1}=c^{-1}[1+r_1 \partial_t - r_2 \nabla^2 + r \partial_{t}^{\kappa} + \cdots].
\end{equation}
Finally using Eqs.~(\ref{V_def}) and~(\ref{fact_prod}), and replacing the
result in Eq.~(\ref{Gandcentralfactor}) we obtain
\begin{multline}
\label{GandV}
G= \mbox{Tr} a(x_2,t_2)\bar{a}(x_1,t_1) \\
\int D\phi D\bar{\phi} \exp\biggl(-\int dt
d^{d}x [\bar{\phi}(\lambda V)^{-1}\phi -a\bar{\phi}-\bar{a}\phi]\biggr).
\end{multline}
Performing the trace operation in Eq.~(\ref{GandV}) and after applying a rescaling of
the fields, we find the effective action of the problem of directed percolation
with incubation times,
\begin{multline}
\label{TLDPaction}
S=\int dt d^{d}x \biggl[ \tilde{\phi}( \partial_{t}^{\kappa}+ \tau \partial_{t} - D_{0}\nabla^{2} + r_{0}) \phi + \frac{1}{2} u_{0} \tilde{\phi}
\phi^{2} \\
- \frac{1}{2} u_{0} \tilde{\phi}^2 \phi \biggr].
\end{multline}
Here $r_0 \propto p_c-p$ and the fields $\phi$ and $\tilde{\phi}$ depend on
space and time. For $\kappa=1$, Eq.~(\ref{TLDPaction}) is reduced to the DP
case. Thus, in our problem we consider $0<\kappa<1$. The generalization of the
Cardy-Sugar formalism, in contrast with the master equation method, does not simply add 
the rates for any possible infection from time $t$
to another time $t'$ in the future, but it properly takes into account the multiple counting of events which may occur if
there is more than one way of reaching a given state at time $t'$.
  
\section{Mean-field approximation}
\label{DPlevyMeanField}
\subsection{Critical exponent $\beta$}
In this section we will find the mean-field values of the critical
exponents of the theory. If fluctuations effects are neglected, the field $\phi$ can be interpreted
as a density field, and consequently above criticality it scales as
\begin{equation}
\phi \propto |p_c-p|^\beta, \qquad p>p_c.
\end{equation}
We start by finding the classical equations of motion. In order
to do so we consider a variation of the action in Eq.~(\ref{TLDPaction}), with respect to the
fields $\phi$ and $\tilde{\phi}$. If we define the Lagrangian density as
\begin{equation}
\label{lagrangian}
\mathcal{L}=\tilde{\phi}( \partial_{t}^{\kappa}+ \tau \partial_{t} - D_{0}\nabla^{2} + r_{0}) \phi + \frac{1}{2} u_{0} \tilde{\phi}
\phi^{2} - \frac{1}{2} u_{0} \tilde{\phi}^2 \phi,
\end{equation}
the corresponding equations of motion are
\begin{equation}
\label{classi1}
\partial_t^{\kappa}\phi+r_0 \phi+\frac{1}{2}u_0(\phi^2-2\tilde{\phi}\phi)-D_0\nabla^2\phi=0,
\end{equation}
\begin{equation}
\label{classi2}
r_0
\tilde{\phi}+\frac{1}{2}u_0(2\tilde{\phi}\phi-\tilde{\phi}^{2})+(-1)^{\kappa}\partial_t^{\kappa}\tilde{\phi}-
D_0\nabla^2\tilde{\phi}=0,
\end{equation}
after using Eq.~(\ref{lagrangian}). A solution where $\tilde{\phi}=0$
would be equivalent to not considering the noise
fluctuations in the Langevin-like equation. From
Eq.~(\ref{classi2}) we see that $\tilde{\phi}=0$ is indeed a classical
solution, as long as Eq.~(\ref{classi1}) is verified:
\begin{equation}
\label{class_12}
 D_0\nabla^2\phi-\partial_t^{\kappa}\phi=r_0 \phi+\frac{1}{2}u_0\phi^2.
\end{equation}
A particular solution of this equation of motion can be
obtained if we neglect the temporal and spatial dependence of the
field $\phi$---that is, a mean-field approximation. Therefore, Eq.~(\ref{class_12}) becomes
\begin{equation}
\label{class_13}
r_0 \phi+\frac{1}{2}u_0\phi^2=0,
\end{equation}
giving two solutions, for $r_0>0$ (below criticality),
\begin{equation}
\phi=0,  \qquad p<p_c,
\end{equation}
and for $r_0<0$ (above criticality),
\begin{equation}
\phi=\frac{-2r_0}{u_0},  \qquad p>p_c.
\end{equation}
Therefore, in a mean-field approximation $\phi \sim r_0$, and from here
we obtain that
\begin{equation}
\label{beta_above_MF}
\beta^{MF}=1.
\end{equation}
%

\subsection{Critical exponents $\nu_\perp$ and $\nu_\parallel$}

In order to calculate the exponents $\nu_\perp$ and $\nu_\parallel$,
we will analyze the scaling behavior of the correlation function
$G^{(1,1)}(x,t)$, around the Gaussian fixed point when the
interaction terms in the action are neglected. Below criticality, we do not expect any longer a
temporal exponential decay of $G^{(1,1)}(x,t)$, as happens in the case of pure DP, but
a power-law behavior. This is due to the fact that the infections can happen at
at very large times. However, the spatial decay of $G^{(1,1)}(x,t)$ is an
exponential decay, since the spatial diffusion is short ranged. Therefore
$G^{(1,1)}(x,t)$ decays exponentially in the limit of large $x$ and as a power
law in the limit of large times. We proceed to
write the Fourier-Laplace transform of $G^{(1,1)}(x,t)$ as
\begin{equation}
\label{F-LG(1,1)}
G^{(1,1)}(x,t) = \int \frac{d^{d}k}{(2\pi)^{d}} e^{ikx}
\int_{\gamma-i\infty}^{\gamma +i\infty} \frac{dE}{2\pi i}\frac{e^{Et}}{E^{\kappa}+D_0
k^2+r_0}.
\end{equation}
If we make the change of variables $E=E^{\prime}r_0^{1/\kappa}$ and
$k=\frac{k^{\prime}}{D_0^{1/2}}r_0^{1/2}$, the correlation function
can be rewritten as follows:
\begin{equation}
\label{F-LG(1,1)_scaling}
G^{(1,1)}(x,t) =  r_0^{1/\kappa}\biggl(\frac{r_0^{1/2}}{D_0^{1/2}}\biggr)^d F\biggl(r_0^{1/\kappa}t, \frac{r_0^{1/2}x}{D_0^{1/2}}\biggr).
\end{equation}
Consequently, time and space scale as $t\sim r_0^{-1/\kappa}$ and
$x\sim r_0^{-1/2}$, respectively. At criticality $r_0=0$, and therefore
any temporal and spatial scale is divergent. We can then define
the critical exponents $\nu_\perp$, which describes how the spatial
correlation length diverges at criticality, the exponent
$\nu_\parallel$, describing the divergent behavior of the temporal
correlation length, and the dynamic exponent
$z=\nu_\parallel/\nu_\perp$, such that
\begin{equation}
\label{MF_exponents}
\nu_\perp^{MF}=\frac{1}{2}, \qquad \nu_\parallel^{MF}=\frac{1}{\kappa},
\qquad z^{MF}=\frac{2}{\kappa}.
\end{equation}
The value of these exponents are given at a mean-field level, as we
have derived them neglecting the interactions in the action in order
to compute $G^{(1,1)}(x,t)$.

Next, we will find how $G^{(1,1)}(x,t)$ decays below
criticality. We should notice that the Laplace transform involved in Eq.~(\ref{F-LG(1,1)}) cannot be solved exactly.
Thus, in Appendix~\ref{appendix0} we compute how this integral behaves asymptotically in
the limit of $t\to \infty$, finding
\begin{equation}
G^{(1,1)}(x,t) \stackrel{t\to \infty}{\sim} \frac{1}{t^{1+\kappa}} \qquad(p<p_c).
\end{equation}
We should compare this result with the DP problem, where $G^{(1,1)}(x,t)$
decays exponentially below $p_c$. At criticality we can perform similar
calculations setting $x\to \infty$ (see Appendix~\ref{appendix0}), and obtain
\begin{equation}
G^{(1,1)}(\infty,t) \stackrel{t\to \infty}{\sim} \frac{1}{t^{1-\kappa}} \qquad (p=p_c).
\end{equation}
Consequently we find that $G^{(1,1)}(x,t)$ behaves with
different power laws at criticality and below criticality. At
criticality, $G^{(1,1)}$ follows a power-law decay given by an
exponent:
\begin{equation}
\label{TLDP_MFdelta}
\delta^{MF}=1-\kappa.
\end{equation}
In this way we have shown that at a mean-field level, the
critical behavior can be described by continuously varying exponents
with the L\'evy parameter $\kappa$. For $\kappa=1$, we recover the DP exponents.

\section{Field-theoretical analysis}
\label{DPlevyFieldTheory}

\begin{figure*}
\centering
\includegraphics[width=120mm]{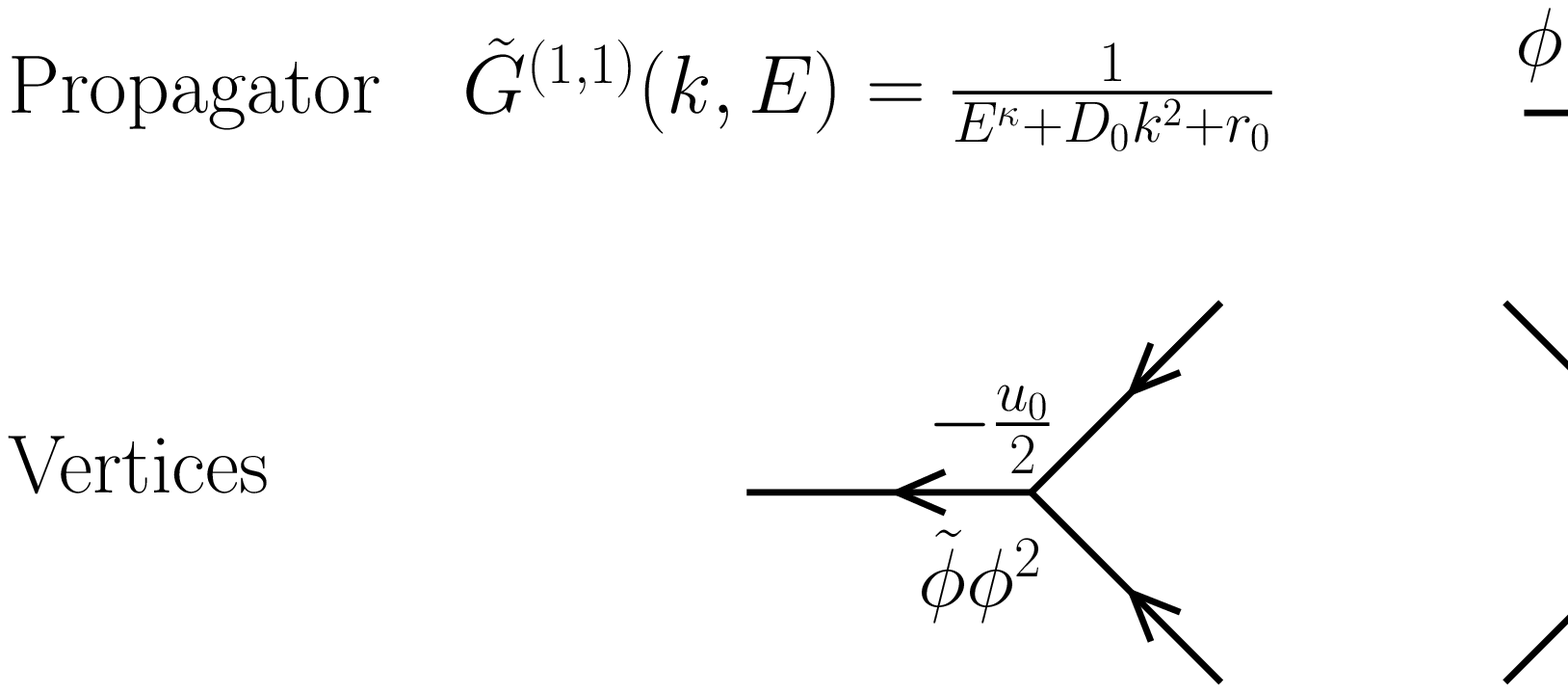}
\caption{
\label{FIG_TLDPfeynmrules}
Propagator and vertices for the problem of directed percolation with
incubation times.
}
\end{figure*}

In this section we will include the effect of fluctuations in our analysis, and therefore
a mean-field approach cannot be considered any longer. Instead we will apply field-theoretical techniques
which will allow us to perform the calculation of the critical exponents below
the upper critical dimension. We start by computing the canonical dimensions for
the various quantities  appearing in the action in
Eq.~(\ref{TLDPaction}), simply by considering the dimensionless
nature of the action. In addition, the time-reversal symmetry $(\tilde{\phi} \rightarrow
-\phi ;\phi  \rightarrow -\tilde{\phi})$, still valid in this problem,
suggests the use of equal canonical dimensions for both fields, $\tilde {\phi}$
and $\phi$. Therefore,
\begin{equation} 
[\tilde{\phi}]=[\phi]=\omega^{(1-\kappa)/2} k^{d/2},
\end{equation}
and the dimensions of the fields depend on the L\'evy parameter $\kappa$.
The canonical dimensions of the diffusion constant $D_{0}$ and the coupling
constant $u_0$ are
\begin{equation}
[D_{0}]= \omega^\kappa k^{-2}
\end{equation}
and
\begin{equation}
[u_0]=\omega^{(3\kappa-1)/2} k^{-d/2},
\end{equation}
respectively. In order to calculate the upper critical dimension $d_c$, we express
the canonical dimension of the coupling constant in terms of momentum units only, as follows
\begin{equation}
\label{u_0_dimensions}
\biggl[\frac{ u_0^2}{D_{0}^{(3\kappa-1)/\kappa}}\biggr]= k^{(6\kappa-2)/\kappa-d}.
\end{equation}
Hence, we see from Eq.~(\ref{u_0_dimensions}) that the coupling constant becomes dimensionless at the value of the
upper critical dimension $d_c$,
\begin{equation}
\label{TLDPdc}
d_c=\frac{6\kappa-2}{\kappa},
\end{equation}
below which the fluctuation effects become important.
We should notice that Eq.~(\ref{TLDPdc}) gives a negative $d_c$ when
$\kappa\leq 1/3$. For these values of $\kappa$, a mean-field
theory rather than a field-theoretical approach should be 
implemented. Consequently, in this section we only consider $1/3<\kappa<1$.

The Feynman rules for the propagator and the
vertices of the theory are formulated in Fig.~\ref{FIG_TLDPfeynmrules}. The propagator $G^{(1,1)}(k,E)$ is represented by a straight line,
and its expression can be obtained from the free action, taking the Laplace transform of time, and the Fourier transform of
the spatial dimensions into momentum space, given as
\begin{equation}
\label{TLDPpropagator}
\tilde{G}^{(1,1)}(k,E)=\frac{1}{E^{\kappa}+D_{0}k^{2}+r_{0}}.
\end{equation}
We have neglected in Eq.~(\ref{TLDPpropagator}) the linear term $\tau E$ with respect
 to $E^\kappa$, since in the low-energy limit ($E\to
0$), the latter term is dominant. The main difference with respect to DP
is the modification of the propagator due to the long-range temporal infections,
given by the non-Markovian term $E^{\kappa}$. Notice that the vertex interactions are not
altered with respect to DP.

In what remains of this section we proceed with the renormalization of
the theory. We will apply mass, field, and diffusion constant
renormalizations to absorb the divergences of the two-point vertex
function $\Gamma^{(1,1)}$. The divergences of $\Gamma^{(2,1)}$ will be
considered in the coupling constant renormalization, and finally we will
renormalize the composite operator $(\tilde{\phi}\phi)$. 

\subsection{Mass and field renormalizations}

\begin{figure*}
\centering
\includegraphics[width=110mm]{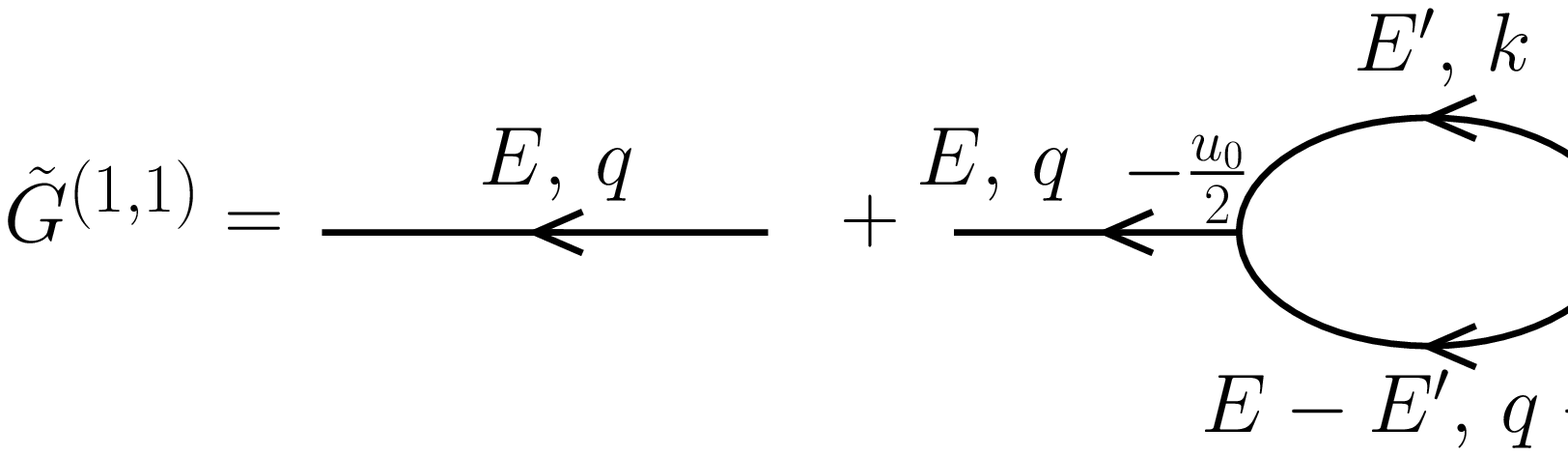}\\
\includegraphics[width=110mm]{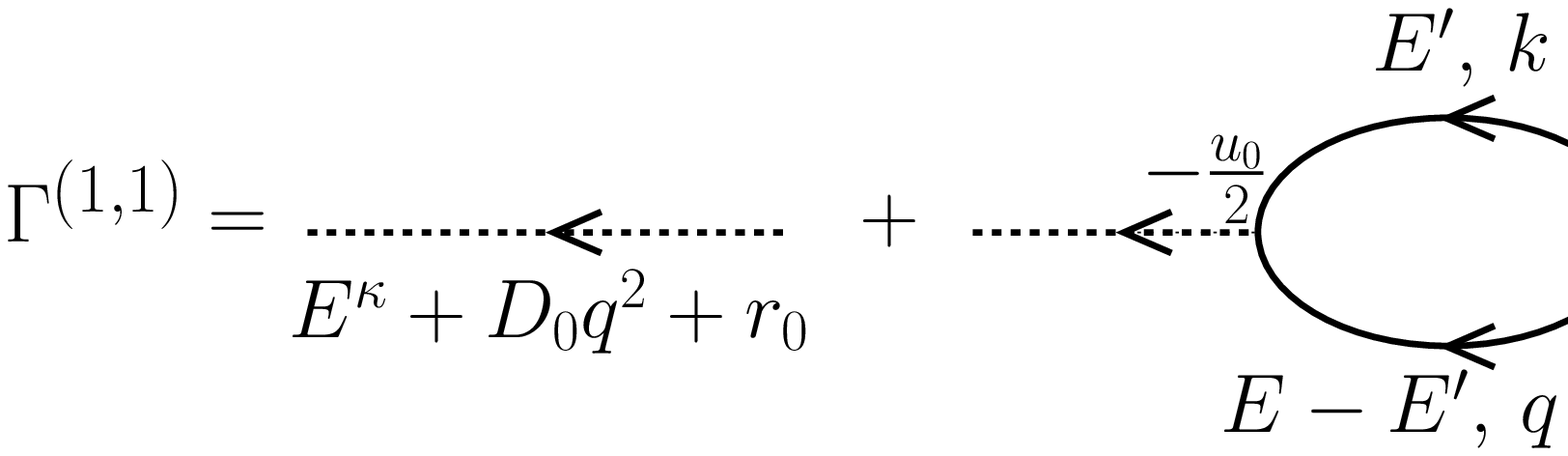}
\caption{
\label{FIG_TLDPGandGamma11}
Feynman diagrams which contribute to the expansions of the propagator $\tilde{G}^{(1,1)}$ and
the two-point vertex function $\Gamma^{(1,1)}$ up to one loop.
}
\end{figure*}

We start by writing the two-point vertex function $\Gamma^{(1,1)}$ at one
loop. Figure~\ref{FIG_TLDPGandGamma11} shows the diagrammatic expansion
up to one loop---that is,
\begin{multline}
\label{TLDPbareGamma11}
\Gamma^{(1,1)}= E^{\kappa}+D_{0}q^{2}+r_{0} + \frac{u_{0}^2}{2} \int
\frac{dE^{\prime}}{(2\pi i)} \int \frac{d^{d}k}{(2\pi)^{d}} 
\\
\times \frac{1}{[E^{\prime \kappa}+D_{0}k^{2}+r_{0}][(E-E^{\prime})^{\kappa}+
D_{0}(q-k)^{2}+r_{0}]}.
\end{multline}
This vertex function has two kinds of divergences. One kind may
happen at $d=4-2/\kappa$, and we assume it is absorbed into a redefinition
of the bare mass $r_0$ to a renormalized mass $r_R$. A second kind of
divergence may happen at $d=6-2/\kappa$, and it will be absorbed into
a renormalization constant of the fields.

\begin{figure}
\centering
\includegraphics[width=70mm]{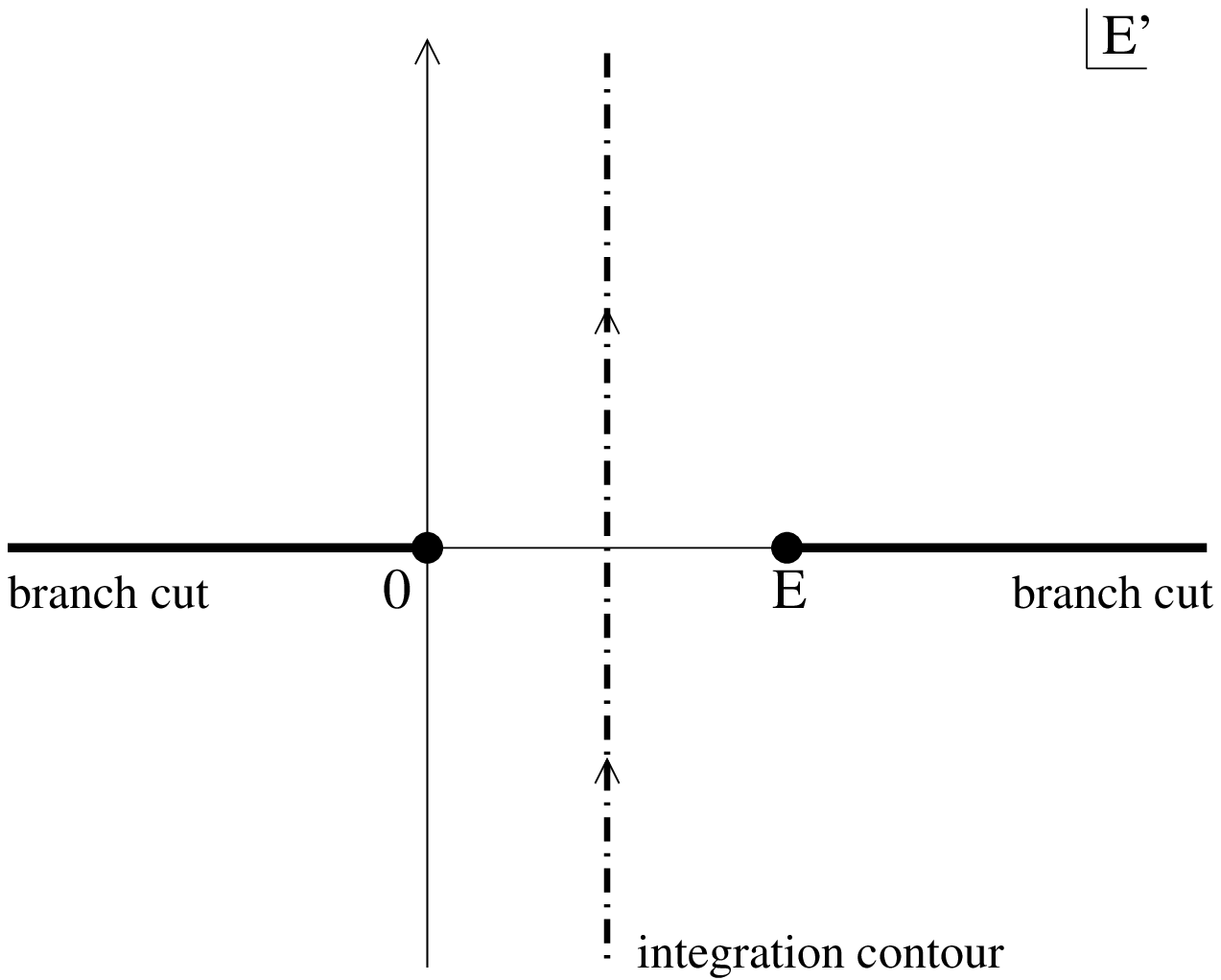}
\caption{
\label{FIG_branchcuts}
Branch points and branch-cut topology for the integral in Eq.~(\ref{TLDP_Zphi}).
}
\end{figure}

We will work in a Laplace-Fourier space constituted by an energy $E$,
considered positive and real, and a
momentum $k$. We define the normalization point (NP) such that the
external energy is evaluated at an arbitrary scale $E=\zeta$, while the
external momentum is set to zero---that is, $q=0$. We define the first
renormalization condition
\begin{equation}
\label{cond1}
\frac{\partial \Gamma^{(1,1)}_{R}}{\partial(E^{\kappa})}\biggl|_{NP}=1.
\end{equation}
The renormalization of the fields defines the renormalization
constants $Z_{\phi}$ and $Z_{\tilde{\phi}}$, such that
\begin{equation}
\phi_R=Z_{\phi}^{-1/2} \phi \, , \qquad \tilde{\phi}_R=Z_{\tilde{\phi}}^{-1/2} \tilde{\phi}.
\end{equation}
Nevertheless, due to the time-reversal symmetry, we can choose $Z_{\phi}=Z_{\tilde{\phi}}$.
The two-point vertex function, calculated by cutting off the external
propagators to the correlation function $\tilde{G}^{(1,1)}$, is then
\begin{equation}
\Gamma^{(1,1)}_{R}= \biggl(\tilde{G}^{(1,1)}_{R}\biggr)^{-1}=  Z_\phi \Gamma^{(1,1)}.
\end{equation}
Inserting this into the renormalization condition, Eq.~(\ref{cond1}),
we obtain the expression for the field renormalization constant,
\begin{equation}
\label{TLDP_Zphi}
Z_{\phi}^{-1}=\frac{\partial \Gamma^{(1,1)}}{\partial(E^\kappa)}\biggl|_{NP}.
\end{equation}
Unfortunately, we were unable to evaluate the integral involved in the expression of
$\Gamma^{(1,1)}$ in Eq.~(\ref{TLDPbareGamma11}) exactly, and consequently we
must rely on an analysis of its asymptotic behavior at the singular
points. Applying standard complex variable theory, we can see that the
integral in Eq.~(\ref{TLDPbareGamma11}) presents two logarithmic branch points: if we
write $E^{\prime \kappa}=e^{\kappa \ln E^{\prime}}$, we identify
one branch point in $E^{\prime}=0$. In the same way it is possible to
see that $E^{\prime}=E$ is the second branch point. There are no poles
in the first Riemann sheet, and therefore we consider the branch-cut
topology shown in Fig.~\ref{FIG_branchcuts}.

We will replace the bare mass $r_0$ by the renormalized mass $r_R$,
assuming that the mass renormalization is already done. For simplicity, when
evaluating the integrals in Eq.~(\ref{TLDP_Zphi}), we will set $r_R=0$ and $q=0$. We start by performing the momentum integration
which can be solved exactly in a straightforward manner, giving
\begin{equation}
\label{Gamma11after_momemtum}
\Gamma^{(1,1)}\biggl|_{NP} = E^{\kappa}+\frac{u_{0}^2}{2} S_d
\frac{\pi}{2} \csc\biggl(\frac{d \pi}{2}\biggr) \frac{1}{D_0^{d/2}}
E^{(1-\epsilon/2)\kappa } I \biggl|_{E=\zeta},
\end{equation}
with
\begin{equation}
\label{I}
I=\int_{1/2-i\infty}^{1/2+i\infty} \frac{du}{(2\pi i)}
\frac{u^{\kappa (d/2-1)}-(1-u)^{\kappa (d/2-1)}}{
(1-u)^{\kappa}-u^{ \kappa}}.
\end{equation}
In Eq.~(\ref{Gamma11after_momemtum}), $\epsilon=d_c-d$ and $S_d=2
\pi^{d/2}/\Gamma(d/2)$. We have made the change of variable $E^{\prime}=E u$, where
$E$ is real and positive. Taking the derivative with respect to $E^{\kappa}$
and evaluating in the NP we obtain the expression for the field renormalization
constant $Z_\phi$:
\begin{equation}
\label{Z_E_integration}
Z_{\phi}^{-1} = 1 + \frac{u_{0}^2}{4} S_d \csc\biggl(\frac{d \pi}{2}\biggr) \frac{\pi}{D_0^{d/2}}
\biggl(1-\frac{\epsilon}{2}\biggr) \zeta^{-\kappa \epsilon/2} I.
\end{equation}
In order to solve the integral $I$, we notice that there are two branch points present, one at $u=0$
and another at $u=1$. We were unable to evaluate this integral exactly,
and consequently in Appendix~\ref{appendix1-field_ren}
we analyze the asymptotic behavior of the integral at the
possible points where divergences may occur. We find that there is no other divergence
present in Eq.~(\ref{Z_E_integration}), except for the one reabsorbed in the definition of the
renormalized mass $r_R$. The direct consequence of this is that 
the field renormalization coefficient remains constant---that is,
$Z_\phi=1+\mathrm{const}$---and under a suitable rescaling of the fields it is
possible to redefine $Z_\phi$ such that at one loop order we have
\begin{equation}
Z_\phi=1.
\end{equation}
In conclusion, according to one-loop calculations, the field renormalization is not required in the theory and
\begin{equation}
\Gamma_R^{(1,1)}\biggr|_{NP}=\Gamma^{(1,1)}\biggr|_{NP},
\end{equation}
which proves that the bare propagator is the full propagator for our theory.

\subsection{\label{subsec_diff} Diffusion constant renormalization}

\begin{figure*}
\centering
\includegraphics[width=140mm]{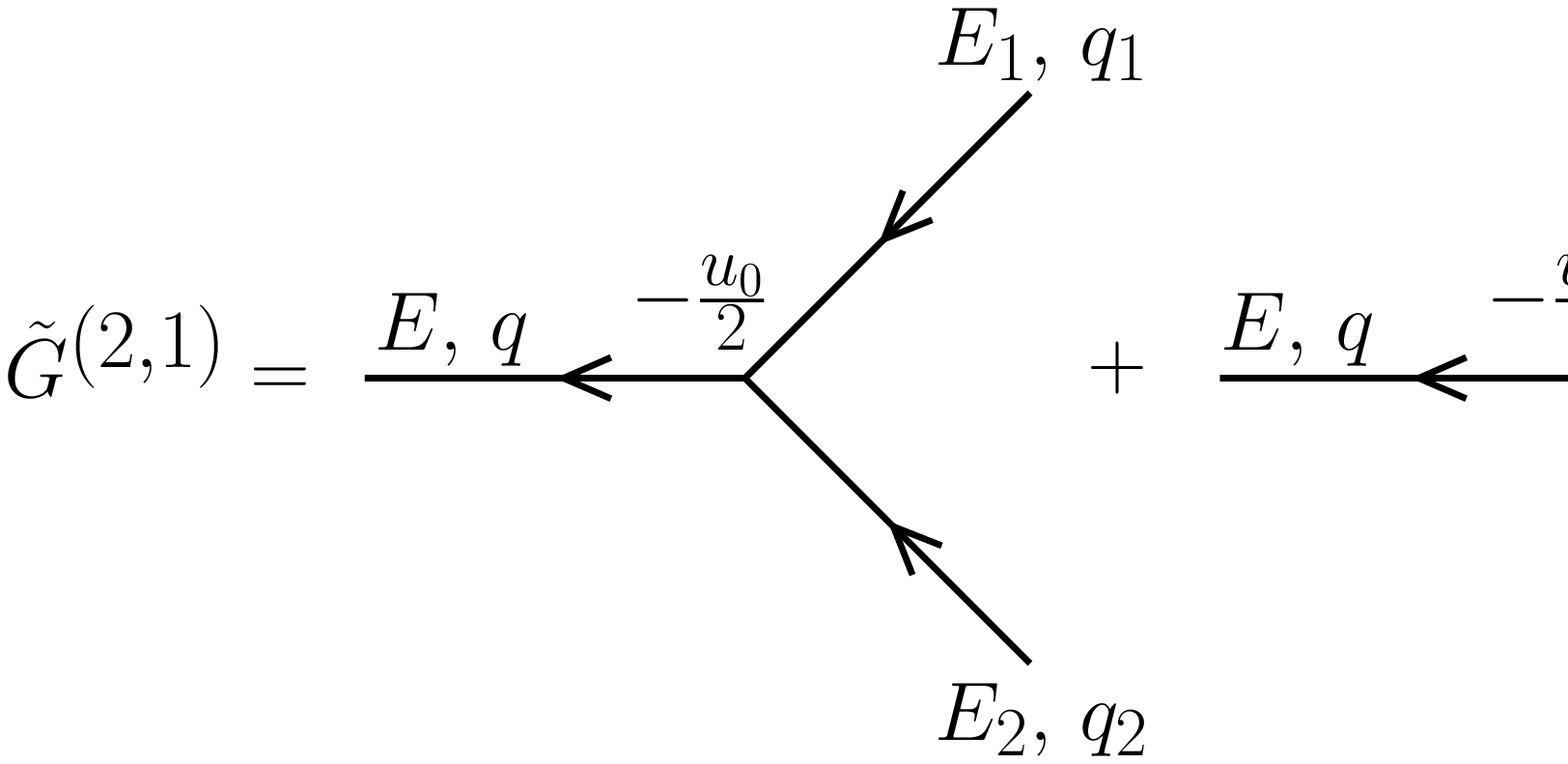}\\
 \includegraphics[width=140mm]{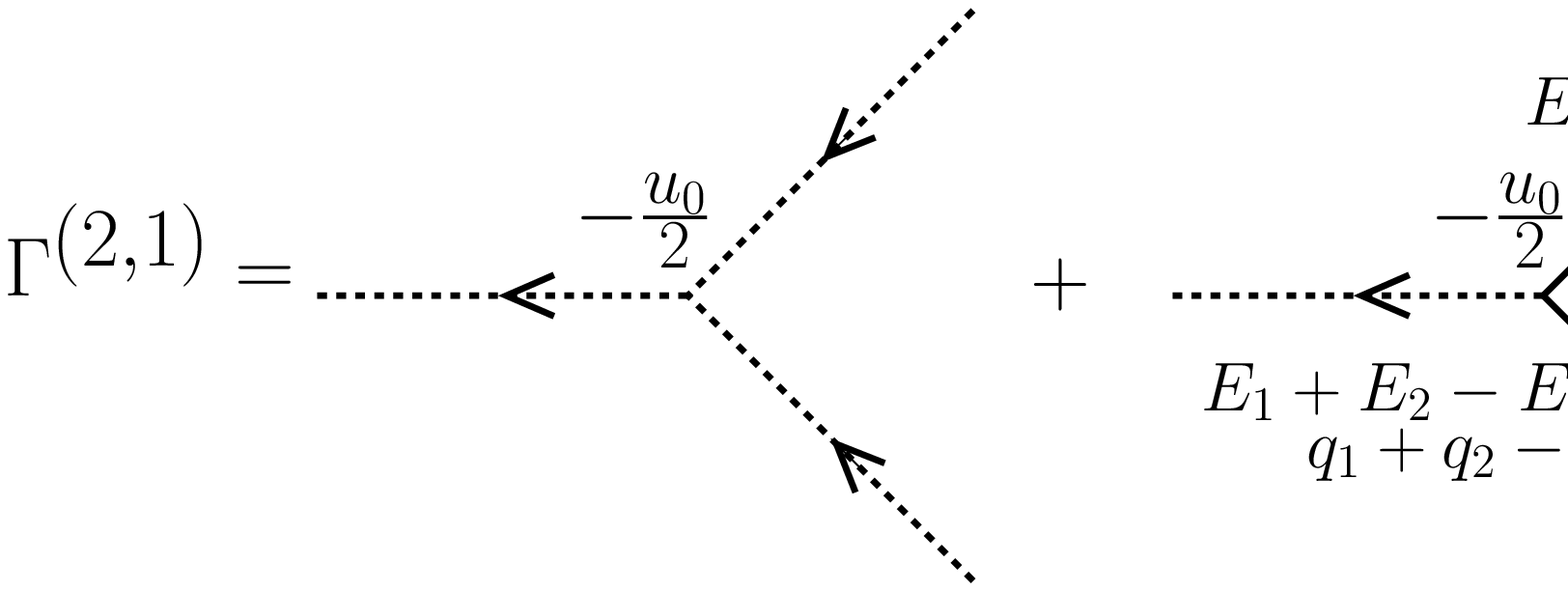}
\caption{
\label{FIG_TLDPGandGamma21}
Perturbative expansions for the correlation function $\tilde{G}^{(2,1)}$ and
the corresponding vertex function $\Gamma^{(2,1)}$ up to one loop.
}
\end{figure*}

In the case of the diffusion constant renormalization we proceed in a similar manner to
that described in the previous subsection, since in this case we also have the technical
difficulty that  the integrals involved in the renormalization cannot be
calculated exactly. Then, we analyze the asymptotic behavior of the integrals
at the different singularities in order to determine the divergences.

We impose the renormalization condition
\begin{equation}
\label{TLDP_D_renor}
\frac{\partial\Gamma^{(1,1)}_{R}}{\partial
q^{2}}\biggl|_{NP}=D_{R}\equiv Z_{D}D_{0},
\end{equation}
which defines the renormalized diffusion constant $D_R$ and the
renormalization constant $Z_D$:
\begin{equation}
Z_{D}=D_{0}^{-1} \frac{\partial\Gamma^{(1,1)}}{\partial q^{2}}\biggl|_{NP}.
\end{equation}
We simplify our calculations setting from the beginning
in our analysis $r_R=0$. After performing the momentum integration we obtain
\begin{multline}
\label{Gamma_IE1_IE2}
\frac{\partial\Gamma^{(1,1)}}{\partial
q^{2}}\biggl|_{E=\zeta}=D_0+\frac{u_0^2}{2}\frac{S_d}{4(2\pi)^{d}
D_{0}^{d/2-1}}\pi \csc(\frac{d\pi}{2})\\
\times \biggl[-I_{E 1}+\frac{1}{d}I_{E 2}\biggr], 
\end{multline}
where the integrals $I_{E 1}$ and $I_{E 2}$ are given in
Appendix~\ref{appendix1-diff_ren} by Eqs.~(\ref{I_E1}) and~(\ref{I_E2}), respectively. In Appendix~\ref{appendix1-diff_ren} we obtain
the divergences of these integrals through an analysis of
the their asymptotic behavior at the singular points. At the upper
critical dimension, we find logarithmic divergences. The integrals cannot be
solved analytically, and therefore we calculate the coefficients of the
divergences. The results are given by Eqs.~(\ref{I_E1_asymp}) and~(\ref{I_E2_asymp}). Inserting these results into Eq.~(\ref{Gamma_IE1_IE2}), the expression for $D_R$ is given by 
\begin{multline}
\label{TLDP_D_R_expan}
D_{R}=D_{0}
Z_{D}=D_0+\frac{u_0^2}{D_{0}^{d/2-1}}\frac{S_d}{2^{\kappa(d/2-3)+5}(2\pi)^{d}} \\
\times \csc\biggl(\frac{d\pi}{2}\biggr)\frac{\zeta^{-\kappa\epsilon/2}}{\kappa\epsilon}
[F_{1}(\kappa)+F_{2}(\kappa)], \qquad \epsilon \to 0,
\end{multline}
where the functions $F_{1}(\kappa)$ and $F_{2}(\kappa)$ are
\begin{equation}
F_{1}(\kappa)=2 \biggl(2-\frac{1}{\kappa}\biggr)\frac{\sin(\pi\kappa)}{\sin^2(\frac{\pi}{2}\kappa)}
\end{equation}
and
\begin{equation}
F_{2}(\kappa)=\frac{1}{\kappa\sin^{3}(\frac{\pi}{2}\kappa)}\biggl[\cos\biggl(\frac{\pi}{2}\kappa\biggr)+\frac{4\kappa^2-3\kappa+1}{3\kappa-1}\cos\biggl(\frac{3\pi}{2}\kappa\biggr)\biggr].
\end{equation}
Although the approximations we have used to compute the
behavior of the integrals are rather drastic, they lead one to obtain the coefficients of
the divergences in a transparent way.

\subsection{Coupling constant renormalization}

Turning now to the coupling constant renormalization, we wish to
construct and follow an equivalent procedure to tackle the integrations
that will appear in this part of the renormalization process. We start
by defining the renormalized coupling constant $u_R$ as
\begin{equation}
\label{TLDPu_Rcond}
u_{R} = - \Gamma^{(2,1)}_{R}\biggl|_{NP} = - Z_\phi^{3/2}
\Gamma^{(2,1)}\biggl|_{NP}= -\Gamma^{(2,1)}\biggl|_{NP}.
\end{equation}
We should notice that in Eq.~(\ref{TLDPu_Rcond}) we have used the fact
that there is no field
renormalization and therefore substituted $Z_\phi=1$. In Fig.~\ref{FIG_TLDPGandGamma21} we show the Feynman
diagrams contributing to $\Gamma^{(2,1)}$ up to one loop.

The dressed vertex function $\Gamma^{(2,1)}$, truncated at two-loop order is, then,
\begin{multline}
\label{TLDP_Gamma21}
\Gamma^{(2,1)}= - u_{0} + u_{0} (-u_{0})^{2} \int \frac{dE^{\prime}}{(2\pi i)}\int \frac{d^{d}k}{(2\pi)^{d}} \\
\times \frac{1}{(E^{\prime \kappa}+D_{0}k^{2})[(E_1-E^{\prime})^{\kappa}+
D_{0}(q_1-k)^{2}]} \\ \times \frac{1}{[(E_1+E_2-E^{\prime})^{\kappa}+
D_{0}(q_1+q_2-k)^{2}]}
\\
+\mathrm{permutation}(E_1,q_1 \leftrightarrow E_2,q_2). 
\end{multline}
The second Feynman diagram contribution, represented by a
$\mathrm{permutation}(E_1,q_1 \leftrightarrow E_2,q_2)$, is an
integration with value equal to the
first integration in Eq.~(\ref{TLDP_Gamma21}), but with the external energy $E_1$ and momentum $q_1$ exchanged with
$E_2$, and $q_2$, respectively. In order to make the computation of the integral in Eq.~(\ref{TLDP_Gamma21}) easier, we
choose the renormalization point NR as
$q_1=q_2=0$, $E_2=0$, and $E_1=\zeta$. This choice is arbitrary and should
not modify the final results. Thus, evaluating Eq.~(\ref{TLDP_Gamma21}) in the normalization
point, we have
\begin{multline}
\label{TLDP_Gamma21_atNP}
\Gamma^{(2,1)}= - u_{0} + 2 u_{0}^{3} \int
\frac{dE^{\prime}}{(2\pi i)} \int \frac{d^{d}k}{(2\pi)^{d}} \\
\times \frac{1}{(E^{\prime \kappa}+D_{0}k^{2})[(E-E^{\prime})^{\kappa}+
D_{0}k^{2}]^{2}}.
\end{multline}
The factor of 2 comes from counting the contribution of the second
Feynman diagram. We proceed by integrating first over the momentum,
and we find
\begin{equation}
\label{TLDP_Gamma21_final}
\Gamma^{(2,1)}= - u_{0} \biggl[1 - \frac{u_{0}^{2}}{D_0^{d/2}} \frac{\pi
\csc(d\pi/2) S_d}{(2\pi)^{d}} I_{E1} \biggr].
\end{equation}
Remarkable, the $I_{E 1}$ is the same integral which first appeared in the
$D_0$ renormalization. We then
make use of the result in Eq.~(\ref{I_E1_asymp}) and
inserting it into Eq.~(\ref{TLDP_Gamma21}) we finally obtain
\begin{equation}
\label{TLDP_u0_expan}
u_R= u_{0} \biggl[1 + \frac{u_{0}^{2}}{D_0^{d/2}} \frac{
\csc(d\pi/2) S_d}{2^{\kappa(d/2-3)+2}(2\pi)^{d}}
F_1(\kappa)\frac{\zeta^{-\kappa\epsilon/2}}{\kappa\epsilon}\biggr], \, \, \epsilon \to 0.
\end{equation}
We are now able to define and calculate the dimensionless
renormalized coupling constant $g_R$. Accordingly to Eq.~(\ref{u_0_dimensions}), it
can be defined as
\begin{equation}
\label{gR_mu}
g_R=\frac{u_R}{D_R^{d/4}} \mu^{-\epsilon/2},
\end{equation}
where $\mu$ is a momentum scale and therefore related to $\zeta$ by
$\mu=\zeta^{\kappa/2}/D_R^{1/2}$. As we have used an energy scale in
the renormalizations, it is convenient to write $g_R$ in terms of
$\zeta$, as follows:
\begin{equation}
\label{gR_zeta}
g_R=\frac{u_R}{D_R^{(d-\epsilon)/4}} \zeta^{-\epsilon \kappa/4}.
\end{equation}
Inserting the expansions of $D_R$ [Eq.~(\ref{TLDP_D_R_expan})] and
$u_R$ [Eq.~(\ref{TLDP_u0_expan})], into Eq.~(\ref{gR_zeta})
we have finally
\begin{multline}
\label{TLDP_gR_expan}
g_R=\frac{u_0}{D_0^{(d-\epsilon)/4}}\zeta^{-\epsilon
\kappa/4}\biggl[1+\biggl(\frac{u_0}{D_0^{d/4}}\zeta^{-\epsilon
\kappa/4}\biggr)^{2} \\
\times \frac{S_d \csc(d \pi/2)}{2^{\kappa(d/2-3)+7}(2\pi)^{d}}\frac{1}{\kappa
\epsilon} F_3(\kappa)\biggr],  \qquad \epsilon \to 0,
\end{multline}
where 
\begin{equation}
F_3(\kappa)=(32-{d_c}) F_1(\kappa)+d_{c} F_{2}(\kappa).
\end{equation}
The next subsection is dedicated to study the final renormalization
step: the composite operator ($\tilde{\phi}\phi$) renormalization.

\subsection{Composite operator renormalization}
%
\begin{figure*}[t]
\centering
\includegraphics[width=110mm]{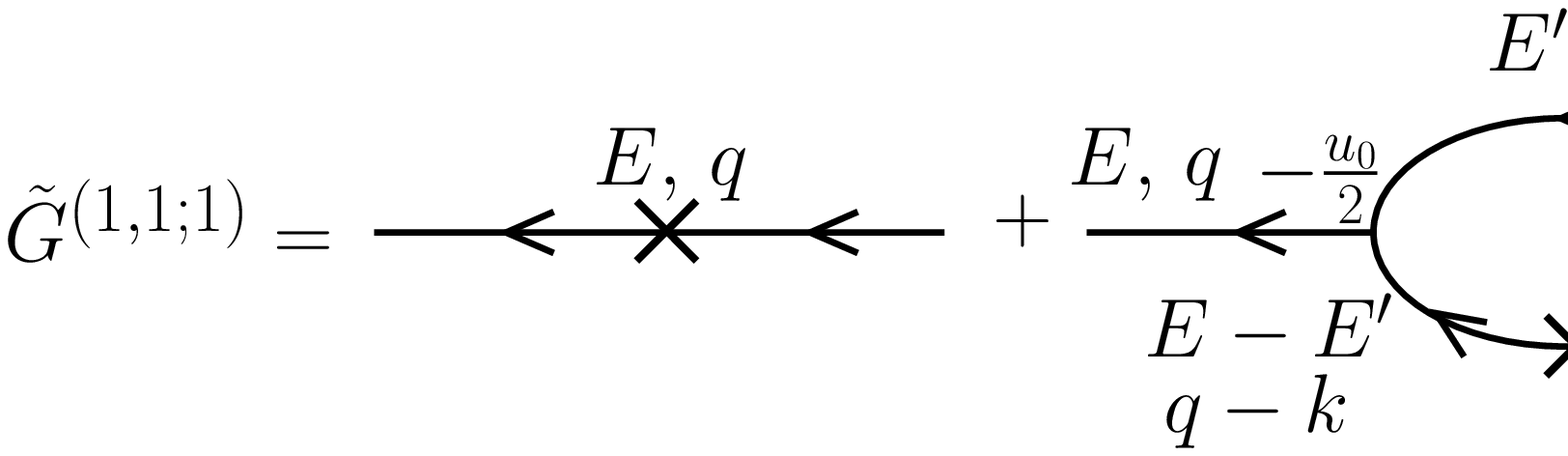}\\
\includegraphics[width=110mm]{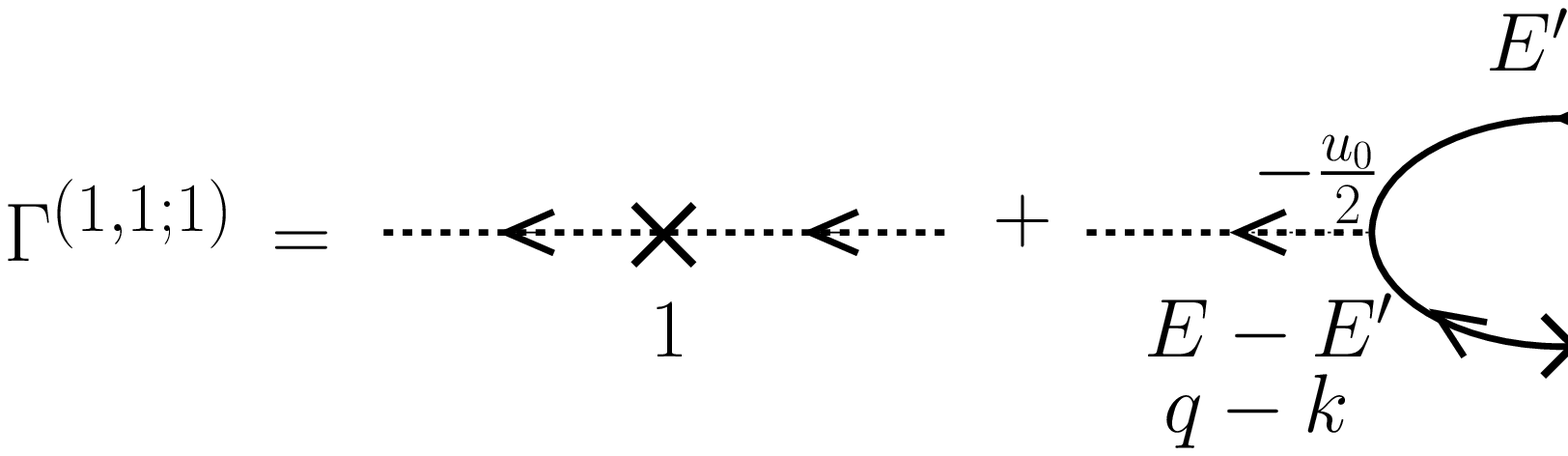}
\caption{
\label{FIG_TLDPcompositeFeynRules}
Expansions of the correlation function $G^{(1,1;1)}$ and
$\Gamma^{(1,1;1)}$ at one loop. The insertion of the composite operator ($\tilde{\phi} \phi$) is represented by a cross.
}
\end{figure*}

We start by defining the renormalized composite operator
$(\tilde{\phi}\phi)_{R}$ as follows:
\begin{equation}
(\tilde{\phi}\phi)_{R}=Z_{\tilde{\phi}\phi}^{-1}(\tilde{\phi}\phi).
\end{equation}
Then the renormalized two-point correlation function with the insertion of
the composite operator can be written as
\begin{equation}
\label{TLDPG111}
\tilde{G}^{(1,1;1)}_{R}=\langle (\tilde{\phi}\phi)_{R} \tilde{\phi}_{R}
\phi_{R}\rangle=Z_{\tilde{\phi}\phi}^{-1} \tilde{G}^{(1,1;1)},
\end{equation}
and the corresponding renormalized vertex function
$\Gamma_{R}^{(1,1;1)}$ is defined by cutting off the external
propagator in Eq.~(\ref{TLDPG111}):
\begin{equation}
\label{TLDPGamma111}
\Gamma_{R}^{(1,1;1)}= [\tilde{G}^{(1,1)}_{R}]^{-1} \tilde{G}^{(1,1;1)}_{R}
[\tilde{G}^{(1,1)}_{R}]^{-1}= Z_{\tilde{\phi}\phi}^{-1}\Gamma^{(1,1;1)}.
\end{equation}
We impose the normalization condition
\begin{equation}
\Gamma_{R}^{(1,1;1)}\biggr|_{NP} \equiv 1,
\end{equation}
and using Eq.~(\ref{TLDPGamma111}) we obtain the expression for the
renormalized constant $Z_{\tilde{\phi}\phi}$,
\begin{equation}
 Z_{\tilde{\phi}\phi}=\Gamma^{(1,1;1)}\biggr|_{NP}.
\end{equation}
The next step is computing the unrenormalized vertex function
$\Gamma^{(1,1;1)}$ evaluated on the renormalization point NP, chosen
as $q=0$ and $E=\zeta$. The Feynman diagrams corresponding to the
dress vertex function $\Gamma^{(1,1;1)}$ up to one loop are
shown in Fig.~\ref{FIG_TLDPcompositeFeynRules}.

The simplest way to calculate
$\Gamma^{(1,1;1)}$ is, first, shifting above criticality where the
renormalized mass $r_R$ is different from its value at criticality $r_{R c}$. We define a parameter $\Delta_0=r_{R}-r_{R c}$
as a measure of the departure from criticality, and we take the
derivative of $\Gamma^{(1,1)}$ with respect to $\Delta_0$, as follows:
\begin{multline}
\Gamma^{(1,1;1)}\biggl|_{NP}=\frac{\partial\Gamma^{(1,1)}}{\partial
\Delta_0}\biggr|_{NP} \\
= 1+\frac{u_{0}^2}{2} \int \frac{dE^{\prime}}{(2\pi i)} \frac{1}{(E-E^{\prime})^{\kappa}-E^{\prime \kappa}} \\
\times \biggl[\int
\frac{d^{d}k}{(2\pi)^{d}}\frac{(-1)}{(E^{\prime\kappa}+D_{0}k^{2}+\Delta_{0})^{2}} \\
+ \int \frac{d^{d}k}{(2\pi)^{d}}\frac{1}{[(E-E^{\prime})^{\kappa}+D_{0}k^{2}+\Delta_{0}]^{2}} \biggr]\biggr|_{E=\zeta}.
\end{multline}
Second, in order to evaluate $\Gamma^{(1,1;1)}$ at criticality we
take the limit $\Delta_0 \to 0$. After performing the momentum
integrations, we have
\begin{equation}
\label{TLDPGamma111expan}
\Gamma^{(1,1;1)}\biggl|_{NP}=1+\frac{u_{0}^2}{D_0^{d/2}}\frac{S_d}{8(2\pi)^{d}}(d-2)\pi
\csc\biggl(\frac{d\pi}{2}\biggr) I_{2},
\end{equation}
where
\begin{equation}
\label{I_2}
I_{2}=\int_{E/2-i\infty}^{E/2+i\infty}
\frac{dE^{\prime}}{(2\pi i)} \frac{E^{\prime
\kappa(d/2-2)}-(E-E^{\prime})^{\kappa(d/2-2)}}{(E-E^{\prime})^{\kappa}-E^{\prime \kappa}}\biggl|_{E=\zeta}.
\end{equation}
In Appendix~\ref{appendix1-composite_ren} we obtain the asymptotic behavior of $I_{2}$ at the
logarithmic divergence [Eq.~(\ref{comp_asimpt})]. Substituting this result into the expression of $\Gamma^{(1,1;1)}$ in
Eq.~(\ref{TLDPGamma111expan}), we have finally
\begin{multline}
Z_{\tilde{\phi}
\phi}=1-\frac{u_{0}^2}{D_0^{d/2}}\frac{S_d}{2^{\kappa(d/2-3)+3}(2\pi)^{d}}(d-2) \\
\times \frac{\sin\biggl[\frac{\pi}{2}\kappa\biggl(\frac{d}{2}-2\biggr)\biggr]}{\sin\biggl(\frac{\pi}{2}\kappa\biggr)}
\frac{\zeta^{-\kappa \epsilon/2}}{\kappa \epsilon}\csc\biggl(\frac{d\pi}{2}\biggr),  \qquad \epsilon \to 0.
\end{multline}
In this way we complete the renormalization procedures
required to absorb any possible divergent term in the vertex
functions up to one loop. In the following section we will write
down the renormalization-group equations and calculate the critical
exponents.

\section{Callan-Symanzik Equation}
\label{DPlevyCallan}
Having performed all the renormalizations required for the theory, we are now able to
calculate the renormalization-group equation for
$\Gamma^{(1,1)}_R$ and $\Gamma^{(1,1;1)}_R$ at criticality and derive
the critical exponents. We will derive the
Callan-Symanzik equations considering a normalization scale $\mu$ in units
of momentum and make the change to $\zeta$ through the relation $\mu
\frac{\partial}{\partial
\mu}=\frac{2}{\kappa}\zeta\frac{\partial}{\partial \zeta}$. Subsequently the $\mu$ dependence will disappear when we
express it in terms of physical quantities, such as energy and momentum.
We start by using the fact that
the unrenormalized $\Gamma^{(1,1)}$ does not depend on the
normalization scale introduced by the normalization point NP, having
then
\begin{eqnarray}
\label{renorEq}
\biggl( \mu \frac{d}{d\mu}\biggr)_{u_0,D_0} \biggl [ Z_{\phi}^{-1}\Gamma^{(1,1)}_{R}(\mu,D_R,g_R)\biggr] & = & 0, 
\end{eqnarray}
and replacing now the total derivative with partial derivatives, we find
\begin{equation}
\label{TLDPCallam-Symanzik}
\biggl[\mu \frac{\partial}{\partial\mu}- \gamma_{\phi}+\gamma_{D}D_{R}\frac{\partial}{\partial
D_{R}}+\beta(g_{R})\frac{\partial}{\partial g_{R}}\biggr]\Gamma^{(1,1)}_{R}(\mu,D_R,u_R)=0,
\end{equation}
where the flow equations, the beta and gamma
functions, are define as follows: 
\begin{equation}
\beta(g_{R})= \mu \biggl(\frac{\partial g_{R}}{\partial\mu}\biggr)_{u_0,D_0}, \nonumber
\end{equation}
\begin{equation}
\gamma_{\phi}= \mu \biggl(\frac{\partial \ln
  Z_{\phi}}{\partial\mu}\biggr)_{u_0,D_0}, \qquad 
\gamma_{D} = \mu \biggl(\frac{\partial \ln Z_{D}}{\partial\mu}\biggr)_{u_0,D_0}.
\end{equation}
The beta function can be calculated using Eq~(\ref{TLDP_gR_expan}),
and it reads as 
\begin{multline}
\label{TLDPfirst_beta}
\beta(g_{R}) =\frac{2}{\kappa}\zeta \biggl(\frac{\partial g_{R}}{\partial\zeta}\biggr)_{u_0,D_0} = -\frac{\epsilon}{2}\biggl(\frac{u_0}{D_0^{(d-\epsilon)/4}}\zeta^{-\epsilon
\kappa/4}  \biggr) \\
+ \frac{3 b}{2 \kappa}\biggl(\frac{u_0}{D_0^{(d-\epsilon)/4}}\zeta^{-\epsilon
\kappa/4}\biggr)^3 +O(u_0^{5}),
\end{multline}
where 
\begin{equation}
\label{b_function}
b(\kappa)=\frac{- S_d
\csc(d\pi/2)}{2^{\kappa(d/2-3)+7}(2\pi)^{d}} F_3(\kappa)
\end{equation}
is a positive and finite function of $\kappa$ in the domain of
interest---that is, $\frac{1}{3}<\kappa < 1$.
Up to first order in $u_0$, $g_R$ can be written as
\begin{equation}
g_R \sim \frac{u_0}{D_0^{(d-\epsilon)/4}}\zeta^{-\epsilon \kappa/4},
\end{equation}
and inserting it into Eq~(\ref{TLDP_gR_expan}) we obtain
\begin{equation}
\frac{u_0}{D_0^{(d-\epsilon)/4}}\zeta^{-\epsilon
\kappa/4}=g_R+\frac{b^3}{\kappa \epsilon} g_R^{3}+O(g_R^{5}).
\end{equation}
Using this result we rewrite Eq.~(\ref{TLDPfirst_beta}) obtaining an
expression of the beta function in terms of $g_R$:
\begin{equation}
\label{TLDP_beta_gR}
\beta(g_{R})=-\frac{\epsilon}{2} g_R+\frac{b}{\kappa}g_R^{3}+O(g_R^{5}),
\end{equation}
which vanishes at the fixed point 
\begin{equation}
g_R^{*}=\sqrt{\frac{\kappa \epsilon}{2 b}}.
\end{equation}
\begin{figure}
\centering
\includegraphics[width=80mm]{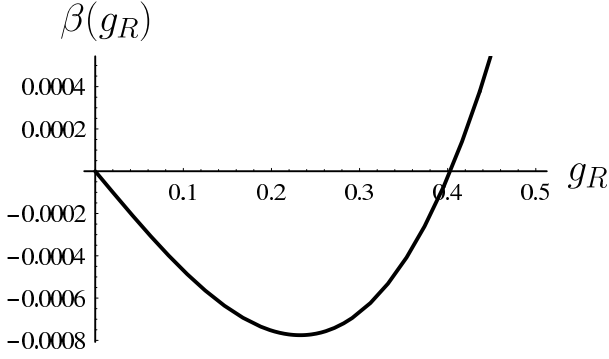}
\caption{
\label{FIG_TLDPbeta}
$\beta$ function $\beta(g_{R})$ for $\kappa=0.7$ and $\epsilon=0.01$.
} 
\end{figure}
The fixed point $g^{*}_R$ is an infrared-stable fixed point as is possible to see from Fig.~\ref{FIG_TLDPbeta}. The
renormalization-group equation~(\ref{TLDPCallam-Symanzik}), evaluated
at $g^{*}_R$, can be written as 
\begin{equation}
\label{TLDPCallam-Symanzik_fixedpoint}
\biggl[\mu \frac{\partial}{\partial\mu}- \gamma^{*}_{\phi}+\gamma^{*}_{D}D_{R}\frac{\partial}{\partial
D_{R}}\biggr]\Gamma^{(1,1)}_{R}(\mu,D_R,u_R)=0,
\end{equation}
where $\gamma^{*}_{\phi}=\gamma_{\phi}(g^{*}_R)=0$ and
$\gamma^{*}_{D}=\gamma_{D}(g^{*}_R)$ at one loop is
\begin{equation}
\gamma^{*}_{D}=\frac{2}{F_3(\kappa)}[F_1(\kappa)+F_2(\kappa)] \epsilon + O(\epsilon^2).
\end{equation}
A solution of Eq.~(\ref{TLDPCallam-Symanzik_fixedpoint}) is given by
\begin{equation}
\label{TLDPGamma_shape}
\Gamma^{(1,1)}_{R}=D_{R} \mu^2 \Phi\biggl(\frac{k}{\mu},\frac{E}{D_{R}^{1/\kappa} k^{2/\kappa}}\biggr).
\end{equation}
In order to write Eq.~(\ref{TLDPCallam-Symanzik_fixedpoint}) in terms of
the physical quantities of
momentum $k$ and energy $E$, we replace the derivatives on $\mu$
and $D_R$ using the following identities:
\begin{equation}
\label{TLDPident_mu}
\mu \frac{\partial}{\partial\mu}=2-k\frac{\partial}{\partial k}- 2
 \frac{1}{\kappa} E \frac{\partial}{\partial E}
\end{equation}
and
\begin{equation}
\label{TLDPident_D}
D_R \frac{\partial}{\partial D_R}=1 - \frac{1}{\kappa} E \frac{\partial}{\partial E},
\end{equation}
which are easy to derive from direct application of the rule of
chain. In this way, we eliminate the $\mu$ dependence in the
Callan-Symanzik equation, rewriting it as
\begin{equation}
\label{TLDPCallanSymanzik_k_E}
\biggl[k\frac{\partial}{\partial k}-(2-\gamma_{\phi}^{*}+
\gamma_{D}^{*})+ \frac{1}{\kappa}(2+\gamma_{D}^{*}) E
\frac{\partial}{\partial E}\biggr]\Gamma^{(1,1)}_R(k,E)=0,
\end{equation}
and applying standard methods to solve it; we obtain
\begin{equation}
\label{TLDPGamma11_solution}
\Gamma^{(1,1)}_{R}(k,E)=k^{2-\gamma_{\phi}^{*}+\gamma_D^{*}} \Phi\biggl(\frac{E}{k^{(2+\gamma_D^{*})/\kappa}}\biggr).
\end{equation}
From this equation we can derive that $E \sim k^{(2+\gamma_D^{*})/\kappa}$, and considering the definition of the dynamic 
exponent $z$, $E \sim t^{-1} \sim k^{z}$, we obtain that
\begin{equation}
\label{TLDPz_critical}
z=\frac{2+\gamma_D^{*}}{\kappa},
\end{equation}
which at one loop gives
\begin{equation}
\label{TLDPz_critical_oneloop}
z=\frac{2}{\kappa}+\frac{2[F_1(\kappa)+F_2(\kappa)]}{\kappa
F_3(\kappa)} \epsilon +O(\epsilon^2).
\end{equation}
This expression is valid at one-loop order in the perturbative
expansion. The next step now is to determine the anomalous
dimension. In order to do so, we calculate the two-point correlation
function for  large times $t\to \infty$:
\begin{align}
\label{TLDP_G11atinfitetime}
G^{(1,1)}(x,\infty) &\sim  \int \frac{d^{d}k}{(2\pi)^{d}}
e^{-ikx}\frac{k^{(2+\gamma_D^{*})/\kappa}}{k^{2-\gamma_{\phi}^{*}+\gamma_D^{*}}}
\nonumber \\
&\sim k^{d-2+\eta_\perp}, \qquad \qquad  p\to p_c.
\end{align}
It is straightforward to find the value of the anomalous dimension
through a simple comparison of the two last lines in
Eq.~(\ref{TLDP_G11atinfitetime}), finding that
$\eta_\perp=\frac{2}{\kappa}+\gamma_{\phi}^{*}+\gamma_{D}^{*}(\frac{1}{\kappa}-1)$.
Using the result obtained for $z$ in Eq.~(\ref{TLDPz_critical}) and
making $\gamma_{\phi}^{*}=0$, we have finally
\begin{equation}
\label{TLDP_eta}
\eta_\perp = z(1-\kappa)+2.
\end{equation}
This new relationship at one-loop order between $\eta_\perp$ and $z$ is a direct
consequence of the absence of field renormalization.
We want to compute now how the density of active particles behaves as
$x \to \infty$---that is, the correlation function $G_R^{(1,1)}(\infty,t)$. Therefore, we turn our attention to
Eq.~(\ref{TLDPGamma11_solution}), where we can see that $k\sim
E^{1/(2+\gamma_D^{*})}$, and as $k \to 0$, the vertex function
$\Gamma_{R}^{(1,1)}$ scales as
\begin{equation}
\label{Gamma_E_behaviour}
\Gamma_{R}^{(1,1)}(0,E) \sim
E^{[\kappa/(2+\gamma^{*}_D)](2-\gamma_{\phi}^{*}+\gamma_D^{*})}.
\end{equation}
It is then easy to obtain the temporal dependence of $G_{R}^{(1,1)}$
at $k=0$, using the behavior of the vertex function given by Eq.~(\ref{Gamma_E_behaviour}):
\begin{equation}
G_R^{(1,1)}(x=\infty,t) \sim \int dE e^{Et} \Gamma_{R}^{(1,1) -1}(0,E) 
\sim t^{-1+\kappa - \frac{\kappa \gamma_{\phi}^{*}}{2+\gamma^{*}_D}},
\end{equation}
which gives a power law decay with the exponent:
\begin{equation}
\delta=1-\kappa +\frac{\kappa \gamma_{\phi}^{*}}{2+\gamma^{*}_D}.
\end{equation}
At one loop, because of the absence of field renormalization in the
theory, $\gamma^{*}_{\phi}=0$ and therefore the value of this exponent
coincides at one loop with its mean-field value found in Eq.~(\ref{TLDP_MFdelta}).
This is an expected result, in the sense that when 
field renormalization is not required, the bare propagator, valid to
describe the  density of active particles at a
mean-field level, is itself the full propagator of the theory when
fluctuations effects are taken into account. 

\subsection{$\beta$, $\nu_\perp$, and $\nu_\parallel$ exponents}

In this subsection we shall investigate the renormalization-group
equation for the vertex function $\Gamma^{(1,1;1)}$ at criticality. The results
of these calculations will let us derive the critical exponents 
$\beta$, $\nu_\perp$, and $\nu_\parallel$. The starting point again is the independence of the unrenormalized  
$\Gamma^{(1,1;1)}$ on the momentum scale $\mu$, which lets us write
\begin{equation}
\biggl( \mu \frac{d}{d \mu}\biggr)_{u_0,D_0} Z_{\tilde{\phi}\phi} Z_{\phi}^{-1}\Gamma^{(1,1;1)}_{R}=0.
\end{equation}
We should notice that the parameter which accounts for the shift of criticality $\Delta_{0}=r_R-r_{R c}$ is taken
equal zero, and we will only use it at some point in the
calculations in order to do dimensional analysis. The Callan-Symanzik equation for $\Gamma^{(1,1;1)}$ reads as follows:
\begin{equation}
\label{TLDPGamma111_Callan}
\biggl(\mu \frac{\partial}{\partial\mu}+\gamma^{*}_{\tilde{\phi}\phi}-\gamma^{*}_{\phi}+\gamma^{*}_{D} D_{R} \frac{\partial}{\partial D_{R}}\biggr)\Gamma^{(1,1;1)}_{R}=0,
\end{equation}
where we have already evaluated the gamma function
$\gamma_{\tilde{\phi}\phi}= \mu \frac{\partial
lnZ_{\tilde{\phi}\phi}}{\partial\mu}$ at the $g_R^{*}$
fixed point. At one loop we have
\begin{equation}
\gamma_{\tilde{\phi}\phi}^{*}= \frac{16 (2\kappa-1)}{\kappa F_3(\kappa)\tan(\frac{\pi}{2}\kappa)}\epsilon+O(\epsilon^2).
\end{equation}
We will maintain explicitly $\gamma^{*}_{\phi}$ in the
equations and make it zero in the end, with the only purpose
of pointing out the direct consequences of the absence of field renormalization at one loop.
Through dimensional analysis, we infer a solution of
Eq.~(\ref{TLDPGamma111_Callan}) as follows
\begin{equation}
\label{TLDPGamma_shape}
\Gamma^{(1,1;1)}_{R}=\Phi\biggl(\frac{k}{\mu},\frac{E}{D_{R}^{1/\kappa} k^{2/\kappa}}\biggr).
\end{equation}
Making use of  the identities 
\begin{equation}
\label{TLDPident_mu}
\mu \frac{\partial}{\partial\mu}=-k\frac{\partial}{\partial k}- 2
 \frac{1}{\kappa} E \frac{\partial}{\partial E}
\end{equation}
and
\begin{equation}
\label{TLDPident_D}
D_R \frac{\partial}{\partial D_R}= - \frac{1}{\kappa} E \frac{\partial}{\partial E},
\end{equation}
we can replace the derivative on $\mu$ and $D_R$ in terms of
derivatives in momentum $k$ and energy $E$, to obtain the
Callan-Symanzik equation at criticality:
\begin{equation}
\biggl(k \frac{\partial}{\partial k}+\gamma^{*}_{\phi} -
\gamma^{*}_{\tilde{\phi}\phi}+\frac{(2+\gamma^{*}_{D})}{\kappa} E
\frac{\partial}{\partial E}\biggr)\Gamma^{(1,1;1)}_{R}=0.
\end{equation}
A solution of this equation is given by
\begin{equation}
\label{TLDPGamma111scaling1}
\Gamma^{(1,1;1)}_{R} =
k^{\gamma^{*}_{\tilde{\phi}\phi}-\gamma^{*}_{\phi}}
\Phi\biggl(\frac{E}{k^{(2+\gamma^{*}_{D})/\kappa}} \biggr).
\end{equation}
Now, we can use the scaling form of the vertex function
$\Gamma^{(1,1)}_{R}$ above criticality---that is, $\Gamma^{(1,1)}_{R} \sim k^{2- \gamma^{*}_{\phi}+\gamma^{*}_{D}}
f(k^{-1/\nu_{\perp}} \Delta_{0},E^{-1/\nu_{\parallel}} \Delta_{0})$---to
obtain in an alternative way the scaling form of $\Gamma^{(1,1;1)}_{R}$:
\begin{eqnarray}
\label{TLDPGamma111scaling2}
\Gamma^{(1,1;1)}_{R} & = & \frac{\partial\Gamma^{(1,1)}_{R}}{\partial\Delta_{0}}\biggl|_{\Delta_{0}=0} \nonumber \\
                       & \sim & k^{2- \gamma^{*}_{\phi}+\gamma^{*}_{D}-\nu_{\perp}^{-1}} .
\end{eqnarray}
This equation in comparison with Eq.~(\ref{TLDPGamma111scaling1}) allows
us to find the value of the exponent $\nu_\perp$ as a function of
$\gamma^{*}_{\tilde{\phi}\phi}$ and $\gamma^{*}_{D}$,
\begin{equation}
\nu_{\perp}=\frac{1}{2-\gamma^{*}_{\tilde{\phi}\phi}+\gamma^{*}_{D}}.
\end{equation}
At one-loop order this equation gives
\begin{equation}
\nu_\perp=\frac{1}{2}+\bigg( \frac{8(2\kappa-1)}{\kappa\tan(\frac{\pi}{2}\kappa)}-F_1(\kappa)-F_2(\kappa)\biggr)
\frac{\epsilon}{2F_3(\kappa)}+O(\epsilon^2).
\end{equation}
In addition, through the definition of $z$, we know that $\nu_\parallel=z \nu_\perp$, and therefore we find
\begin{equation}
\nu_\parallel=\frac{1}{\kappa}\biggl( \frac{2+\gamma_D^{*}}{2-\gamma^{*}_{\tilde{\phi}\phi}+\gamma^{*}_{D}}\biggr),
\end{equation}
which at one loop gives
\begin{equation}
\nu_\parallel=\frac{1}{\kappa}+\frac{8 (2\kappa-1)}{\kappa^2 F_3(\kappa)\tan(\frac{\pi}{2}\kappa)}\epsilon+O(\epsilon^2).
\end{equation}
We calculate now the  $\beta$ exponent above criticality, by writing down
how $G^{(1,1)}(x,t)$ behaves in the limit of large times: 
\begin{align}
G^{(1,1)}(x,\infty) &\sim  \int \frac{d^{d}k}{(2\pi)^{d}} \frac{d \omega}{2\pi}
e^{-ikx} k^{-2+\gamma^{*}_{\phi}-\gamma^{*}_{D}}
g\biggl(\frac{k}{\Delta_{0}^{\nu_\perp}},\frac{\omega}{\Delta_{0}^{\nu_{\parallel}}}\biggr)
\nonumber \\
&\sim  |\Delta_0|^{\nu_\parallel+\nu_\perp
  (d-2+\gamma^{*}_{\phi}-\gamma^{*}_{D})}\sim |\Delta_0|^{\nu_\perp
  (d+\eta_\perp-2)} \nonumber \\
&\sim  |\Delta_0|^{2\beta}.
\end{align}
In this way, above criticality we obtain the same relationship valid
for DP---that is,
\begin{equation}
\label{TLDPbeta}
\beta=\frac{\nu_\perp}{2}(d+\eta_\perp-2).
\end{equation}
We can write $d=d_c-\epsilon$, and thus the value of
the $\beta$ exponent at one loop can be calculated using
Eq.~(\ref{TLDPbeta}), given
\begin{multline}
\label{TLDPbetaoneloop}
\beta= 1+\biggl(\frac{8(2 \kappa -1)}{\tan(\pi\kappa/2)}-[F_1(\kappa)+F_2(\kappa)]
\frac{3\kappa-1}{2} - \frac{\kappa F_3(\kappa)}{4} \biggr) \\
\times \frac{\epsilon}{\kappa F_3(\kappa)}+  O(\epsilon^2).
\end{multline}
Nevertheless, in our theory  we have in addition an extra relationship given by Eq.~(\ref{TLDP_eta}) because
of the absence of field renormalization in the theory. Using both
Eqs.~(\ref{TLDP_eta}) and~(\ref{TLDPbeta}), we obtain, at
one-loop order,
\begin{equation}
\label{TLDPbeta_eq2}
2\beta=\nu_\parallel(1-\kappa)+d\nu_\perp.
\end{equation}
The existence of this relationship makes the exponent $\beta$ dependent on the
value of $\nu_\parallel$ and $\nu_\perp$, and therefore reduces the
independent critical exponents from three to two independent critical
exponents, with respect to the DP theory. Therefore, we find a new scaling relation at one loop for the problem of
DP with incubation times. In the following section we argue
that this result is valid to any loop order in perturbation theory.
%
%
\section{Discussion}
\label{DPlevyDiscussion}
In this paper we have formulated and solved a field theory for a
non-Markovian model of directed percolation with the inclusion of long-range
temporal diffusion, which in a context of epidemics can be interpreted as
incubation times. The incubation times are distributed following a L\'evy
distribution, while the spatial diffusion as well as the
interactions are short ranged. We first draw the attention to the fact that the conventional approach of writing a master
equation, in order to apply later a second-quantized formalism, is not
a convenient way  to find the field-theoretical action. This is mainly
due to the fact that the master equation has an infinite number of terms,
a consequence of the  nondisjoint nature of events for different
times. Instead, we have proposed an extension of a method introduced by Cardy and
Sugar~\cite{CardySugar80}, where we included the details of the long-range
temporal diffusion in the effective lattice determined by the infection vectors.
Following this approach, in a rather simple way we have found the action of the
problem.

Second, we found already at a mean-field level that the critical
exponents vary continuously with the L\'evy parameter, signaling the
existence of a new universality class. We also found at a mean-field level
that the two-point correlation function decays as a power law below
criticality, instead of showing an exponential decay as in DP. This is a
consequence of having infections that can be produced at very large times. We
also found that this power-law decay is different from the one obtained for the 
two-point correlation function above criticality.

Subsequently, including fluctuation effects, we have
renormalized the theory at one loop. We have
calculated the renormalization-group equations and
we have determined the critical exponents at one order
in an $\epsilon$ expansion. The critical exponents vary
continuously with the L\'evy parameter and obey at one loop an extra relationship with respect
to DP, Eq.~(\ref{TLDPbeta_eq2}), which is a direct consequence of the absence of field renormalization.

We argue now that the new relationship, Eq.~(\ref{TLDPbeta_eq2}), is valid to all orders in
perturbation theory, which will be true if the absence of field
renormalization occurs at any loop. The  absence of field
renormalization just by power counting is difficult to see in the
renormalization scheme applied here. This is because terms
proportional to $E^\kappa$ in principle can be generated at any loop.
Therefore in our case it will be necessary to check that the coefficients
of these terms are not divergent at the upper critical dimension. The absence of singular field
renormalization to all orders is more clear in other renormalization-group schemes---for
example, the Wilson method or normalization at nonzero momentum, since in
those cases the relevant Feynman amplitudes are always analytic in the
external energies and momenta. Hence no terms like $E^\kappa$ as $E\to0$
can be generated in loop diagrams, even though they are present in the
bare propagator. The absence of renormalization of such singular terms in
the propagator to all orders has long been known for the case of
long-range ferromagnets~\cite{Sak73}, and has also been recognized for
other variants of DP with long-range spatial interactions~\cite{Janssen99, HHoward99} and with both
long-range spatial and temporal interactions~\cite{AKSH2005}.

Finally, we notice that when we set $\kappa$ equal to 1, we do not recover the DP
hyperscaling relation from Eq.~(\ref{TLDPbeta_eq2}). This is
due to the fact that the validity of Eq.~(\ref{TLDPbeta_eq2}) relies
on the absence of field renormalization. This does not
happen in DP, where field renormalization is necessary to absorb a divergence that appears at the upper
critical dimension $d_c=4$. 

Since this work was completed, a related paper~\cite{AKSH2005} has appeared. This
differs from the present one in that long-range effects in both time and
space are considered. The renormalization of this theory is simpler than
the case considered here, because neither the coefficients of $E^\kappa$
nor of $k^\sigma$ in the bare propagator are renormalized, and hence there
are two additional scaling relations rather than the single one found here.
One of the two scaling  relations obtained in~\cite{AKSH2005} is
equivalent to the scaling relation, Eq.~(\ref{TLDPbeta_eq2}), found in this work.

\acknowledgments
%
The author thanks John Cardy for enlightening discussions and suggestions and
Martin Evans, Martin Howard, and Robin Stinchcombe for useful comments. This work was supported by CONICET (Argentina), the British Council-Fundaci\'on
Antorchas (Argentina), and the ORS Award Scheme (UK).

\appendix

\section{\label{appendix0} Asymptotic behavior of the correlation function
  $G^{(1,1)}(x,t)$}
In this appendix we compute how $G^{(1,1)}(x,t)$, Eq.~(\ref{F-LG(1,1)}), behaves asymptotically in
the limit of $t\to \infty$. The contribution to the integral is due only to the presence
of a branch point in $E=0$, since there are no poles in the first Riemann
sheet. If we call $y=|E|$, then 
\begin{multline}
\label{LG(1,1)}
\int_{\gamma-i\infty}^{\gamma +i\infty} \frac{dE}{2\pi i} \frac{e^{Et}}{E^{\kappa}+D_0
k^2+r_0} = \frac{1}{2\pi i} \int_{0}^{\infty} e^{-yt} \\
\times \biggl[\frac{1}{y^\kappa e^{-i\pi \kappa}+D_0 k^2+r_0}-\frac{1}{y^\kappa e^{i\pi \kappa}+D_0 k^2+r_0}\biggr]dy.
\end{multline}
As $t\to\infty$, we see from Eq.~(\ref{LG(1,1)}) that the leading
behavior of the integral comes from the integration domain for small
$y$. We consider then a series expansion of the integrand as follows:
\begin{multline} 
\frac{1}{y^\kappa e^{-i\pi \kappa}+D_0 k^2+r_0}-\frac{1}{y^\kappa
e^{i\pi \kappa}+D_0 k^2+r_0}\\
= \frac{2i}{(D_0 k^2+r_0)^{2}} \sin(\pi \kappa)y^{\kappa} + O(y^{2\kappa}).
\end{multline}
Inserting this result into Eq.~(\ref{LG(1,1)}), we obtain
\begin{multline}
\label{LG(1,1)second}
\int_{\gamma-i\infty}^{\gamma +i\infty} \frac{dE}{2\pi i}\frac{e^{Et}}{E^{\kappa}+D_0
k^2+r_0} \\
\stackrel{t\to \infty}{\sim}\frac{\sin(\pi \kappa)}{\pi(D_0 k^2+r_0)^2} \frac{1}{t^{1+\kappa}} \Gamma(1+\kappa).
\end{multline}
Therefore, the asymptotic
behavior of $G^{(1,1)}(x,t)$ as $t\to \infty$ is given by
\begin{multline}
G^{(1,1)}(x,t) \\
\sim \frac{\Gamma(1+\kappa)\sin(\pi \kappa)}{\pi}  \frac{1}{t^{1+\kappa}} \int
 \frac{d^{d}k}{(2\pi)^{d}} \frac{e^{ikx}}{(D_0 k^2+r_0)^2}
 \\
\sim \frac{1}{t^{1+\kappa}}, \qquad t \to \infty \qquad(p<p_c).
\end{multline}
A similar analysis can be carried out at criticality, where
$r_0=0$. In this case we set $k=0$, and we have
\begin{multline}
G^{(1,1)}(\infty,t) \sim \frac{1}{2\pi i}\int_0^\infty
e^{-yt}\frac{2i\sin{(\pi \kappa)}}{y^\kappa} dy \\
\sim \frac{\sin(\pi\kappa)}{\pi}\frac{1}{t^{1-\kappa}}
\Gamma(1-\kappa),   \qquad t \to \infty  \qquad (p=p_c).
\end{multline}

\section{\label{appendix1} Asymptotic behavior of the integrals in the
  renormalization calculations}

\subsection{\label{appendix1-field_ren} Integrals involved in the field
  renormalization}
We proceed to analyze the asymptotic behavior of integral $I$ in Eq.~(\ref{I}) at the
values of $u$ where divergences may occur---that is, $u=1/2$ and $u=\frac{1}{2}\pm i \infty$. Where such divergences exist, we
will determine the finite coefficients of the integral. At
$u \to 1/2$, the numerator and denominator of the integrand goes to zero. Expanding them around the point $u=1/2$, we find
that they both go to zero with the same order in $(u-1/2)$.
Therefore the integrand remains finite in the limit $u\to 1/2$, and we infer then that the integral itself does not present a divergence in
the integration domain around $u=1/2$.

The next step is to consider the asymptotic behavior as $u\to
\frac{1}{2}\pm i\infty$. To analyze this limit it is convenient to 
make a change of variables $iy=u-\frac{1}{2}$, and the
integration $I$ in Eq.~(\ref{Z_E_integration}) can be rewritten as follows:
\begin{multline}
\label{theta_y_field_int}
I= 2 \int_{0}^{\infty}\frac{dy}{2\pi} y^{\kappa(d/2-2)} \biggl(1+\frac{1}{4 y^{2}}\biggr)^{(d/2-2)\kappa/2} \\
\times \biggl[\frac{e^{i\theta(d/2-1)\kappa}-e^{-i\theta(d/2-1)\kappa}}{e^{-i\theta\kappa}-e^{i\theta\kappa}}\biggr],
\end{multline}
where $\theta\equiv\theta(y)=\arctan(2|y|)$.  We now integrate by parts
in Eq.~(\ref{theta_y_field_int}) and find
\begin{multline}
\label{f_and_f'}
I=\frac{1}{\pi}\int_{0}^{\infty} dy y^{\kappa(d/2-2)}f(y)=\frac{1}{\pi}\biggl[
\frac{y^{ \kappa(d/2-2)+1}}{\kappa(d/2-2)+1}f(y)\biggl|_{0}^{\infty}
\\
-\int_{0}^{\infty} dy \frac{y^{ \kappa(d/2-2)+1}}{\kappa(d/2-2)+1} f^{\prime}(y)\biggr],
\end{multline}
where $f(y)$ is
\begin{equation}
f(y)=\biggl(1+\frac{1}{4 y^{2}}\biggr)^{(d/2-2)\kappa/2}
\biggl[\frac{e^{i\theta(d/2-1)\kappa}-e^{-i\theta(d/2-1)\kappa}}{e^{-i\theta\kappa}-e^{i\theta\kappa}}\biggr].
\end{equation}
The function $f(y)$ is a finite, going to a constant as $y\to \infty$, and
vanishes as $y\to 0$ quickly enough for the first term on the right-hand side
to be convergent for $\kappa(d/2-2)+1<0$---that is, for
$d<4-\frac{2}{\kappa}$. On the other hand, for
$d\geq 4-\frac{2}{\kappa}$, as $y\to \infty$, the first term on the right-hand side
diverges as $y^{\kappa(d/2-2)+1}$. We have assumed that such a divergence was absorbed into a
renormalized mass $r_R$ in the mass renormalization
procedure: the change of variables
performed in Eq.~(\ref{Gamma11after_momemtum}) scales out of the integral the energy
dependence in the second term of $\Gamma^{(1,1)}$, and therefore the
divergences corresponding to mass renormalization in $\Gamma^{(1,1)}$
and $Z_{\phi}^{-1}$ are the same. Thus, it is only the second term on the right-hand side of
Eq.~(\ref{f_and_f'}) that may diverge at
$d=d_c=6-\frac{2}{\kappa}$. There is already a pole at  $d>4-\frac{2}{\kappa}$ in
this term, and thus in order to identify the next divergence we will extend the value
of this term to $d>4-\frac{2}{\kappa}$, applying analytic continuation.

First of all, we need to determine the shape of $f'(y)$ for large
values of $y$. We start by considering the series expansion of
$\theta=\arctan(2y)=\frac{\pi}{2}-\frac{1}{2y}+O(y^{-3})$ valid for
$y>1/2$. Inserting it in the expression of $f(y)$, we find
\begin{multline}
f(y)=\frac{\sin[\frac{\pi}{2}(\frac{d}{2}-1)\kappa]}{\sin\frac{\pi}{2}\kappa}\biggl[1 \\
+\frac{\kappa}{2}\biggl(\frac{1}{\tan\frac{\pi}{2}\kappa}-\frac{(\frac{d}{2}-1)}{\tan[\frac{\pi}{2}(\frac{d}{2}-1)\kappa]}\biggr)\frac{1}{y}+O\biggl(\frac{1}{y^{2}}\biggr) \biggr].
\end{multline}
It is straightforward to see from here that $f^{\prime}(y)\propto
\frac{1}{y^2}+O(\frac{1}{y^3})$. Then, if we call $\alpha=\kappa(d/2-2)$,
the integration in the right-hand side of Eq.~(\ref{f_and_f'})  behaves as
\begin{equation}
\label{last_field_integral}
\int_{0}^{\infty} y^{\alpha+1} f^{\prime}(y) dy \sim \frac{y^{\alpha}}{\alpha}, \qquad y\to \infty.
\end{equation}
This result suggests that the next pole would be at
$\alpha=0$. Nevertheless, at the upper critical dimension
$d_c=6-2/\kappa$, $\alpha$ is different from zero and takes negative
values for any $\kappa$, with $1/3<\kappa<1$. This means that the integral in Eq.~(\ref{last_field_integral}) is
convergent as $y\to \infty$. We show with this calculation that there is no other divergence present in Eq.~(\ref{I}).

\subsection{\label{appendix1-diff_ren} Integrals involved in the diffusion
  constant renormalization}
In the Sec.~\ref{subsec_diff}, the expression of the renormalization constant $Z_{D}$ contains the integrals
\begin{multline}
\label{I_E1}
I_{E 1}=\int_{\frac{E}{2}-i\infty}^{\frac{E}{2}+i\infty}
\frac{dE^{\prime}}{(2\pi i)}
\biggl[2 E^{\prime \kappa(\frac{d}{2}-1)}+
(E-E^{\prime})^{\kappa (\frac{d}{2}-1)} (d-4) \\
-E^{\prime \kappa} (E-E^{\prime})^{\kappa (\frac{d}{2}-2)}(d-2)\biggr]/ [E^{\prime \kappa}-(E-E^{\prime})^{\kappa}]^{2} \biggl|_{E=\zeta}
\end{multline}
and
\begin{multline}
\label{I_E2}
I_{E 2}=\int_{\frac{E}{2}-i\infty}^{\frac{E}{2}+i\infty}
\frac{dE^{\prime}}{(2\pi i)}\biggl\{ [8 E^{\prime \frac{\kappa d}{2}}+
2 E^{\prime \kappa} (E-E^{\prime})^{\kappa (\frac{d}{2}-1)} \\
\times (d-4) d - E^{\prime 2\kappa} (E-E^{\prime})^{\kappa (\frac{d}{2}-2)}(d-2) d \\
-(E-E^{\prime})^{\frac{\kappa d}{2}}(8-6d+d^2)]/[E^{\prime \kappa}-(E-E^{\prime})^{\kappa}]^{3}\biggr\}\biggl|_{E=\zeta}.
\end{multline}
Let us begin to analyze the existence of divergences in $I_{E 1}$. We make a change of variables $E^{\prime}=Eu$, where $E$ is real and
positive. Inserting this change into Eqs.~(\ref{I_E1}) and~(\ref{I_E2}), and after evaluating in
the normalization point $E=\zeta$, it is possible to see that the only divergences in the integrands could appear 
in the limits  $u\to \frac{1}{2}\pm i\infty$. In order to analyze these limits, we make the change of variables $iy=u-\frac{1}{2}$
and $I_{E 1}$  and $I_{E 2}$ look as follows:
\begin{multline}
I_{E 1}= -\frac{\zeta^{-\kappa\frac{\epsilon}{2}}}{4\pi}\int_{0}^{\infty}dy
 y^{\kappa(\frac{d}{2}-3)}\biggl(1+\frac{1}{4y^2}\biggr)^{\frac{\kappa}{2}(\frac{d}{2}-3)}\\
\times \biggl[\frac{(d-2)\{\cos[\theta\kappa(\frac{d}{2}-1)]-\cos[\theta\kappa(\frac{d}{2}-3)]\}}{\sin^{2}(\theta\kappa)}\biggr]
\end{multline}
and
\begin{multline}
I_{E 2}= \frac{\zeta^{-\kappa\frac{\epsilon}{2}}}{8\pi}\int_{0}^{\infty}dy
 y^{\kappa(\frac{d}{2}-3)}\biggl(1+\frac{1}{4y^2}\biggr)^{\frac{\kappa}{2}(\frac{d}{2}-3)} \\
\times \biggl\{2(d-4)d
 \sin\biggl[\theta\kappa\biggl(\frac{d}{2}-2\biggr)\biggr]-(d-2)d\sin\biggl[\theta\kappa\biggl(\frac{d}{2}-4\biggr)\biggr]\\
-(16-6d+d^2)\sin\biggl(\theta \kappa\frac{d}{2}\biggr)\biggr\}/ \sin^{3}(\theta\kappa),
\end{multline}
respectively, being $\theta\equiv\theta(y)=\arctan(2|y|)$. 
Therefore as $y\to\infty$ the integrals diverge as
$y^{\kappa(d/2-3)}$, and at the upper critical dimension
these divergences become logarithmic. We cannot solve the integrals
analytically, and for this reason we only determine the
coefficients of such divergences. We start by expressing the entire
integrand as a function of $\theta$ using
$y=\frac{\tan\theta}{2}$ and  $1+\frac{1}{4y^2}=(\sin
\theta)^{-2}$. For instance, $I_{E 1}$ now looks
\begin{multline}
\label{I_E1_theta}
I_{E 1} = -\frac{\zeta^{-\kappa\frac{\epsilon}{2}}}{4\pi}
\frac{1}{2^{\kappa(\frac{d}{2}-3)+1}} \int_{0}^{\frac{\pi}{2}} d\theta
f(\theta)\\
= -\frac{\zeta^{-\kappa\frac{\epsilon}{2}}}{4\pi}
\frac{1}{2^{\kappa(\frac{d}{2}-3)+1}} \int_{0}^{\frac{\pi}{2}} d\theta
\frac{1}{(\cos\theta)^{\kappa(\frac{d}{2}-3)+2}} \\
\times \frac{(d-2)\{\cos[\theta\kappa(\frac{d}{2}-1)]-\cos[\theta\kappa(\frac{d}{2}-3)]\}}{\sin^{2}(\theta\kappa)}.
\end{multline}
The limit of interest is $y \to \infty$ or, equivalently, $\theta\to \pi/2$.
Therefore, to proceed with our analysis, we can write $\theta=\pi/2-\alpha$, where $\alpha$ is a
variable which tends to zero. Substituting $\theta$ as a function of
$\alpha$ in $I_{E 1}$ in Eq.~(\ref{I_E1_theta}) and taking the limit $\alpha
\to  0$, we have
\begin{multline}
\int_{0}^{\frac{\pi}{2}} d\theta
f(\theta) = \int_{0}^{\frac{\pi}{2}}
d\alpha
\frac{1}{[\cos(\frac{\pi}{2}-\alpha)]^{\kappa(\frac{d}{2}-3)+2}} \\
\times
\frac{(d-2)\{\cos[(\frac{\pi}{2}-\alpha)\kappa(\frac{d}{2}-1)]-
\cos[(\frac{\pi}{2}-\alpha)\kappa(\frac{d}{2}-3)]\}}{\sin^{2}[(\frac{\pi}{2}-\alpha)\kappa]}\\
\stackrel{\alpha \to 0}{\longrightarrow} \frac{(d-2)\{\cos[\frac{\pi}{2}\kappa(\frac{d}{2}-1)]-
\cos[\frac{\pi}{2}\kappa(\frac{d}{2}-3)]\}}{\sin^{2}(\frac{\pi}{2}\kappa)} \\
\times \int_{0}^{\Delta}d\alpha \frac{1}{\alpha^{\kappa(\frac{d}{2}-3)+2}} + \mathrm{finite},
\end{multline}
where $0<\Delta<\pi/2$. We can write  $\kappa(\frac{d}{2}-3)+2=1-\epsilon/2$ with
$\epsilon=d_c-d$, and  then we see that as $\alpha \to 0$, the
integrand $f$ in function of $\alpha$ diverges as
\begin{equation}
f(\alpha) \sim a \alpha^{\frac{\kappa\epsilon}{2}-1}, \qquad \alpha \to 0,
\end{equation}
with $a=\frac{(d-2)\{\cos[\frac{\pi}{2}\kappa(\frac{d}{2}-1)]-
\cos[\frac{\pi}{2}\kappa(\frac{d}{2}-3)]\}}{\sin^{2}(\frac{\pi}{2}\kappa)}$,
a constant number. If we now add and substract this divergence in the expression of
$f(\alpha)$, we can obtain the coefficient of the logarithmic divergence as follows:
\begin{align}
\int_{0}^{\frac{\pi}{2}} d\alpha f(\alpha) &= \int_{0}^{\frac{\pi}{2}}
d\alpha [f(\alpha)-a \alpha^{\frac{\kappa\epsilon}{2}-1}] +
a\int_{0}^{\frac{\pi}{2}}d\alpha \alpha^{\frac{\kappa\epsilon}{2}-1} \nonumber \\
&\sim a \, \,\frac{2}{\kappa\epsilon}, \qquad \qquad \epsilon \to 0.
\end{align}
Inserting this result into Eq.~(\ref{I_E1_theta}), we then have 
\begin{multline}
\label{I_E1_asymp}
I_{E 1} \sim \frac{-1}{\pi
2^{\kappa(\frac{d}{2}-3)+2}}
\frac{\zeta^{-\kappa\frac{\epsilon}{2}}}{\kappa\epsilon}\\
\times \biggl[\frac{(d-2)\{\cos[\frac{\pi}{2}\kappa(\frac{d}{2}-1)]-
\cos[\frac{\pi}{2}\kappa(\frac{d}{2}-3)]\}}{\sin^{2}(\frac{\pi}{2}\kappa)}
\biggr], \, \, \,\epsilon \to 0.
\end{multline}
Preceding in the same way for $I_{E 2}$ we have
\begin{multline}
\label{I_E2_asymp}
I_{E 2} \sim \frac{1}{\pi 2^{\kappa(\frac{d}{2}-3)+3}}
\frac{\zeta^{-\kappa\frac{\epsilon}{2}}}{\kappa\epsilon} \biggl[2(d-4)d
 \sin\biggl[\frac{\pi}{2}\kappa\biggl(\frac{d}{2}-2\biggr)\biggr]\\
-(d-2)d \sin\biggl[\frac{\pi}{2}\kappa\biggl(\frac{d}{2}-4\biggr)\biggr]\\
-(16-6d+d^2)\sin\biggl(\kappa\frac{d\pi}{4}\biggr)\biggr] /\sin^{3}\biggl(\frac{\pi}{2}\kappa\biggr), \qquad \epsilon \to 0.
\end{multline}
\subsection{\label{appendix1-composite_ren} Integrals involved in the composite
  operator renormalization}
The integration in $E^{\prime}$ in Eq.~(\ref{I_2}) can be studied starting by doing the
change of variable $E^{\prime}=E u$, as we did previously, considering
$E$ a real and positive number:
\begin{equation}
\label{TLDPu_int}
I_{2}=\zeta^{\kappa(\frac{d}{2}-3)+1}\int_{\frac{1}{2}-i\infty}^{\frac{1}{2}+i\infty} \frac{du}{(2\pi i)}
\frac{u^{\kappa(\frac{d}{2}-2)}-(1-u)^{\kappa(\frac{d}{2}-2)}}{(1-u)^{\kappa}-u^{\kappa}}.
\end{equation}
The integrand is not divergent as $u\to \frac{1}{2}$, but goes to a
constant as one can see from a Taylor
expansion of the integrand. We can infer then that the possible divergences
may have their origin in the limits $u\to\frac{1}{2}\pm
i \infty$. For the purpose of studying the behavior of the integral
in those limits, we rewrite the integration in Eq.~(\ref{TLDPu_int}), making the
change of variables $iy=u-1/2$, as follows:
\begin{multline}
\label{TLDPy_int}
\int_{\frac{1}{2}-i\infty}^{\frac{1}{2}+i\infty} \frac{du}{(2\pi i)}
\frac{u^{\kappa(\frac{d}{2}-2)}-(1-u)^{\kappa(\frac{d}{2}-2)}}{(1-u)^{\kappa}-u^{\kappa}} \\
= 2 \int_{0}^{\infty}\frac{dy}{2\pi} y^{\kappa(d/2-3)} \biggl(1+\frac{1}{4
y^{2}}\biggr)^{(d/2-3)\kappa/2} \\
\times \biggl[\frac{e^{i\theta(d/2-2)\kappa}-e^{-i\theta(d/2-2)\kappa}}{e^{-i\theta\kappa}-e^{i\theta\kappa}}\biggr].
\end{multline}
In the limit $y\to \infty$, the integrand diverges as
$y^{\kappa(d/2-3)}$, and at the upper critical dimension this
integral become logarithmic divergent. We analyze then this limit by
making use of the relations $y=\frac{\tan\theta}{2}$ and setting
$\theta=\pi/2-\alpha$, such that $\alpha \to 0$. We express
Eq.~(\ref{TLDPy_int}) as a function of $\alpha$:
\begin{multline}
2 \int_{0}^{\infty}\frac{dy}{2\pi} y^{\kappa(d/2-3)} (1+\frac{1}{4 y^{2}})^{(d/2-3)\kappa/2} \\
\times \biggl[\frac{e^{i\theta(d/2-2)\kappa}-e^{-i\theta(d/2-2)\kappa}}{e^{-i\theta\kappa}-e^{i\theta\kappa}}\biggr]
\stackrel{\alpha \to 0}{\longrightarrow} \frac{-1}{\pi 2^{\kappa(\frac{d}{2}-3)+1}}\\
\times \frac{\sin[\frac{\pi}{2}\kappa(\frac{d}{2}-2)]}{\sin[\frac{\pi}{2}\kappa]}
\int_{0}^{\Delta} d\alpha \frac{1}{\alpha^{1-\kappa \epsilon/2}}+\mathrm{finite}.
\end{multline}
The integrand diverges as $\alpha^{\kappa \epsilon/2-1}$ in the
limits of $\alpha
\to 0$ and $\epsilon \to 0$. In order to identify the coefficient of
this divergence, we add and substract the divergence itself from Eq.~(\ref{TLDPy_int}), finding
\begin{multline}
\label{comp_asimpt}
\int_{\frac{1}{2}-i\infty}^{\frac{1}{2}+i\infty} \frac{du}{(2\pi i)}
\frac{u^{\kappa(\frac{d}{2}-2)}-(1-u)^{\kappa(\frac{d}{2}-2)}}{(1-u)^{\kappa}-u^{\kappa}}\\
\sim a \frac{2}{\kappa \epsilon} + \mathrm{finite},  \qquad \epsilon \to 0,
\end{multline}
where $a$ is a constant equal to $a=\frac{-1}{\pi 2^{\kappa(\frac{d}{2}-3)+1}}
\frac{\sin[\frac{\pi}{2}\kappa(\frac{d}{2}-2)]}{\sin[\frac{\pi}{2}\kappa]}$.


\end{document}